\documentclass[aps, twocolumn, pre, superscriptaddress, floatfix]{revtex4-1}
\usepackage{graphicx}
\usepackage{dcolumn}
\usepackage{bm, bbm, eufrak}
\usepackage{amsmath}

\usepackage{bm,graphicx,hyperref}
\hypersetup{%
  breaklinks = {true},
  citecolor = {blue},
  colorlinks = {true},
  linkcolor = {red},
}

\newcommand{\mst}{\ensuremath{m_{\rm st}}}

\def\*#1{\mathbf{#1}}

\begin{document}

\title{Curvature as an external field in mechanical antiferromagnets}

\author{Abigail Plummer}
\thanks{These authors contributed equally to this work.\\ aplummer@princeton.edu, paul.hanakata@gmail.com}
\affiliation{Department of Physics, Harvard University, Cambridge, Massachusetts 02138}
\affiliation{Princeton Center for Complex Materials, Princeton University, Princeton, New Jersey 08540}

\author{Paul~Z.~Hanakata}
\thanks{These authors contributed equally to this work.\\ aplummer@princeton.edu, paul.hanakata@gmail.com}
\affiliation{Department of Physics, Harvard University, Cambridge, Massachusetts 02138}

\author{David~R.~Nelson}
\affiliation{Department of Physics, Harvard University, Cambridge, Massachusetts 02138}

\date{\today}
\begin{abstract}
A puckered sheet is a freestanding crystalline membrane with an embedded array of bistable buckled units. Recent work has shown that the bistable units behave like spins in a two-dimensional compressible Ising antiferromagnet with, however, a coupling to flexural phonons. At finite temperature, this purely mechanical system displays Ising-like phase transitions, which drive anomalous thermal expansion. Here, we show that geometry can be used to control phase behavior: curvature produces a radius-dependent ``external field'' that encourages alignment between neighboring ``spins," disrupting the ordered checkerboard ground state of antialigned neighbors. The effective field strength scales as the inverse of the radius of curvature. We identify this effective field theoretically with both a discrete real space model and a nonlinear continuum elastic model.  We then present molecular dynamics simulations of puckered sheets in cylindrical geometries at zero and finite temperature, probing the influence of curvature on the stability of configurations and phase transitions. Our work demonstrates how curvature and temperature can be used to design and operate a responsive and tunable metamaterial at either the macroscale or nanoscale.

  
\end{abstract}

\maketitle
\section{Introduction}
Mechanical systems composed of coupled bistable units have been explored in recent years for applications in soft robotics, shape memory, and information processing \cite{seffen2006mechanical, oppenheimer-PRE-92-052401-2015, liu2022snap, faber-advancedScience-7-2001955-2020, risso2022highly, udani2021programmable, kang2014complex, chen2021reprogrammable, han-nature-456-898-2008, waitukaitis-PRL-114-055503-2015, lechenault2015generic, shohat2022memory, lindeman2021multiple, jules2022delicate, bense2021complex, yasuda2021mechanical, ruiz2016stm, garcia2022buckling}. An appealing feature of these metamaterials is their tunability---each of $N$ bistable units can be individually inverted, possibly leading to $\sim 2^N$ metastable states and diverse macroscopic behaviors \cite{silverberg2014using}. Tunable materials are of interest for many technological applications, from optical filtering~\cite{cui2010metamaterials, zheludev2012metamaterials} to reconfigurable structures~\cite{zirbel2013accommodating, overvelde2017rational} in which it is desirable to have a single material serve multiple functions. A shared challenge of many tunable materials is determining how to easily and reversibly control microscopic configurations, enabling the desired macroscopic transformations.

\begin{figure*}
\includegraphics[width=17.4cm]{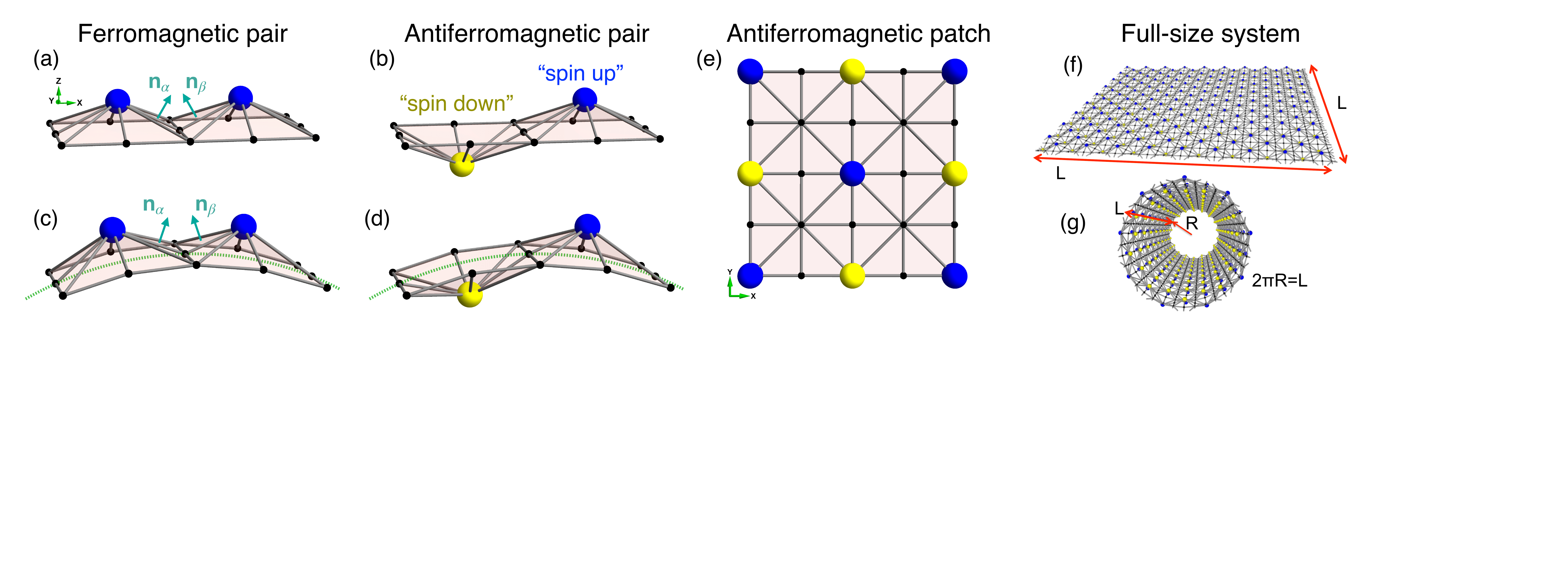}
\caption{Curvature influences the interaction between neighboring buckled dilations.  Dilations are drawn as large (blue or yellow) spheres and undilated host lattice sites are drawn as smaller black spheres. (a) Two dilations buckled in the same direction will have a bending energy contribution which we model as $\hat{\kappa}(1-\mathbf{n}_{\alpha} \cdot \mathbf{n}_\beta )$. (b) Antialigned puckers pay a smaller bending energy penalty, as the plaquettes between the two dilations are parallel. When a background curvature (dotted green line) is introduced, as might happen on a cylinder, it becomes somewhat less costly for puckers to be outwardly aligned  (c), and somewhat more costly to be antialigned (d). (e) A top down view of the network of harmonic springs connecting dilation and host lattice nodes in a square array. (f) An example of a system with 18 $\times$ 18 dilations with planar periodic boundary conditions. (g) The same system as in (f) now rolled into a cylinder to produce an extrinsic radius of curvature $R$.}
\label{fig:schematics}
\end{figure*}

Recently, we proposed that one such system, a free-standing elastic sheet with an array of buckled bistable units, can be understood as a mechanical analog of a compressible Ising antiferromagnet with spin-flexural phonon coupling~\cite{hanakata2022anomalous}. In this system, bistable puckers are created by locally dilating the surface at a regular array of lattice sites embedded in a crystalline membrane---when the dilation is sufficiently large, it becomes energetically favorable for the affected site to buckle, either up or down, into the third dimension. Each buckled dilation acts like a ``spin," and an interaction between neighboring spins is generated via the difference in the elastic energy of different deformation patterns (as in Figs. \ref{fig:schematics}(a,b)). At zero temperature, the energy of a system with stress-free in-plane periodic boundaries is minimized by a checkerboard configuration of up and down puckers, equivalent to an antiferromagnetic spin configuration (Figs. \ref{fig:schematics}b,e ~\ref{fig:mapping}a,b)~\cite{plummer2020buckling, hanakata2022anomalous}. Zero-temperature puckered sheets provide a theoretically tractable system to explore shape memory and metastability, and are relevant to recent experimental realizations of macroscale metasheets~\cite{liu2022snap, faber-advancedScience-7-2001955-2020, risso2022highly}.

When the temperature is increased, thermal energy becomes comparable to the energy barrier between the up and down puckered states, and ``spins" are able to flip. At a critical temperature, the staggered magnetization, which quantifies the checkerboard spin order, drops abruptly and the susceptibility and specific heat diverge. In addition to these standard signatures of an Ising phase transition (albeit with an unusual specific heat exponent~\cite{hanakata2022anomalous}), one finds an anomalous, diverging coefficient of thermal expansion at the critical temperature due to the competition between spin degrees of freedom and out-of-plane thermal fluctuations~\cite{hanakata2022anomalous}. Thermally activated dilation arrays are relevant to experimentally realized puckered atomically thin monolayers such as SnO~\cite{seixas-PRL-116-206803-2016, pacheco2019evolution, daeneke2017wafer}. 

Given an Ising-like mechanical model, a natural next question is: Can we define a mechanical analog of an external field that acts on our ``spins"? An effective external field would ideally enable us to control microscopic spin configurations by varying a macroscopic quantity, allowing a puckered sheet to function as a programmable metamaterial with tunable properties. 

In this paper, we demonstrate with theory and simulations that the extrinsic curvature of the host lattice plays the role of an external field in our system, encouraging dilations to defy antiferromagnet nearest-neighbor coupling and buckle in the same direction. A large curvature corresponds to a high effective uniform magnetic field. Figure \ref{fig:schematics} provides an intuitive understanding of why host lattice curvature can bias dilations to buckle away from the center of curvature---the angle between two aligned or ``ferromagnetic" puckers is smoothed by the presence of curvature, decreasing the cost of bending. Our framework is consistent with observations in the literature of mechanical Ising-like systems with free boundaries adopting curved configurations when nodes are assigned to be in the same state \cite{oppenheimer-PRE-92-052401-2015,seffen2006mechanical, liu2022snap, risso2022highly, faber-advancedScience-7-2001955-2020}. Extrinsic curvature is an appealing candidate stimulus for many applications, as it can often be tuned at the boundary \cite{stein2022efficient}. 


In order to study curvature in a controlled manner, we focus on square arrays of dilations rolled into cylinders (Fig. \ref{fig:schematics}g). This geometry allows us to explore the effect of a background of nonzero mean curvature without the stretching associated with nonzero Gaussian curvature \cite{do2016differential, vella2019buffering}. A cylindrical geometry also allows us to connect more closely to the literature on functionalized carbon nanotubes with defects~\cite{lee2006cycloaddition, goldsmith2007conductance}, van der Waals nanotubes (e.g., MoS$_2$ monolayer wrapped into a cylinder)~\cite{liu2021photoluminescence, cui2018rolling} and ferromagnetic nanotubes~\cite{bachmann2007ordered, buchter2013reversal}. 

We support our claim that a background curvature acts as a biasing external field in Ising-like puckered cylinders with two complementary theoretical models as well as molecular dynamics simulations at both zero and finite temperature. In Sec. \ref{sec:model}, we introduce a computational model for an array of buckled bistable nodes on a cylinder and briefly summarize key simulation results. In Sec. \ref{sec:discrete}, we provide a discrete real space theory based on approximations to the energy used in simulations, and show that couplings between neighboring spins and between spins and curvature take the same forms as terms in the microscopic Ising Hamiltonian. In Sec. \ref{sec:continuum}, we develop a nonlinear continuum model using shallow shell theory, which we use to derive a Landau-like expansion of the energy with a fieldlike coupling between curvature and magnetization. In Sec. \ref{sec:simulations}, we use molecular dynamics simulations to confirm that ferromagnetic buckling is preferred for high curvature and antiferromagnetic buckling is preferred for intermediate or vanishing curvature at zero temperature. We then increase the temperature and track the phase behavior of the system. Finite-size effects are inevitable, since for our cylindrical geometry, we cannot increase the curvature without decreasing the cylinder circumference. Nonetheless, we generate an approximate phase diagram in the curvature-temperature plane showing that finite staggered magnetization can only be maintained at either low temperatures or small curvatures. We conclude by discussing prospects for future work, including studying the influence of higher order couplings to curvature predicted by our theory, arrays of contractile inclusions, systems with nonzero Gaussian curvature, and high-temperature crumpling. 

\section{Model}\label{sec:model}
In this section, we present the computational model used in simulations (Sec. \ref{sec:simulations}) whose behavior we seek to understand using theory (Secs. \ref{sec:discrete} and \ref{sec:continuum}).

To model an antiferromagnetic array of dilations embedded in a thin elastic sheet, we use the energy functional introduced in ref. \cite{seung-PRA-38-1005-1988}, but generalized to have a square microstructure \cite{plummer2020buckling}. Lattice sites are connected by harmonic springs, shown as grey lines in Fig.~\ref{fig:schematics}e, and each triangular plaquette is assigned a normal vector that is used to penalize bending, as shown in Fig.~\ref{fig:schematics}a,c. This type of model has been used to study the mechanics and thermal behavior of atomically thin materials such as graphene and MoS$_2$~\cite{zhang-JMPS-67-2-2014, wan2017thermal, bowick-PRB-95-104109-2017, hanakata-EML-44-101270-2020, morshedifard-JMPS-149-104296-2021}. 

\begin{figure*}
\includegraphics[width=0.9\textwidth]{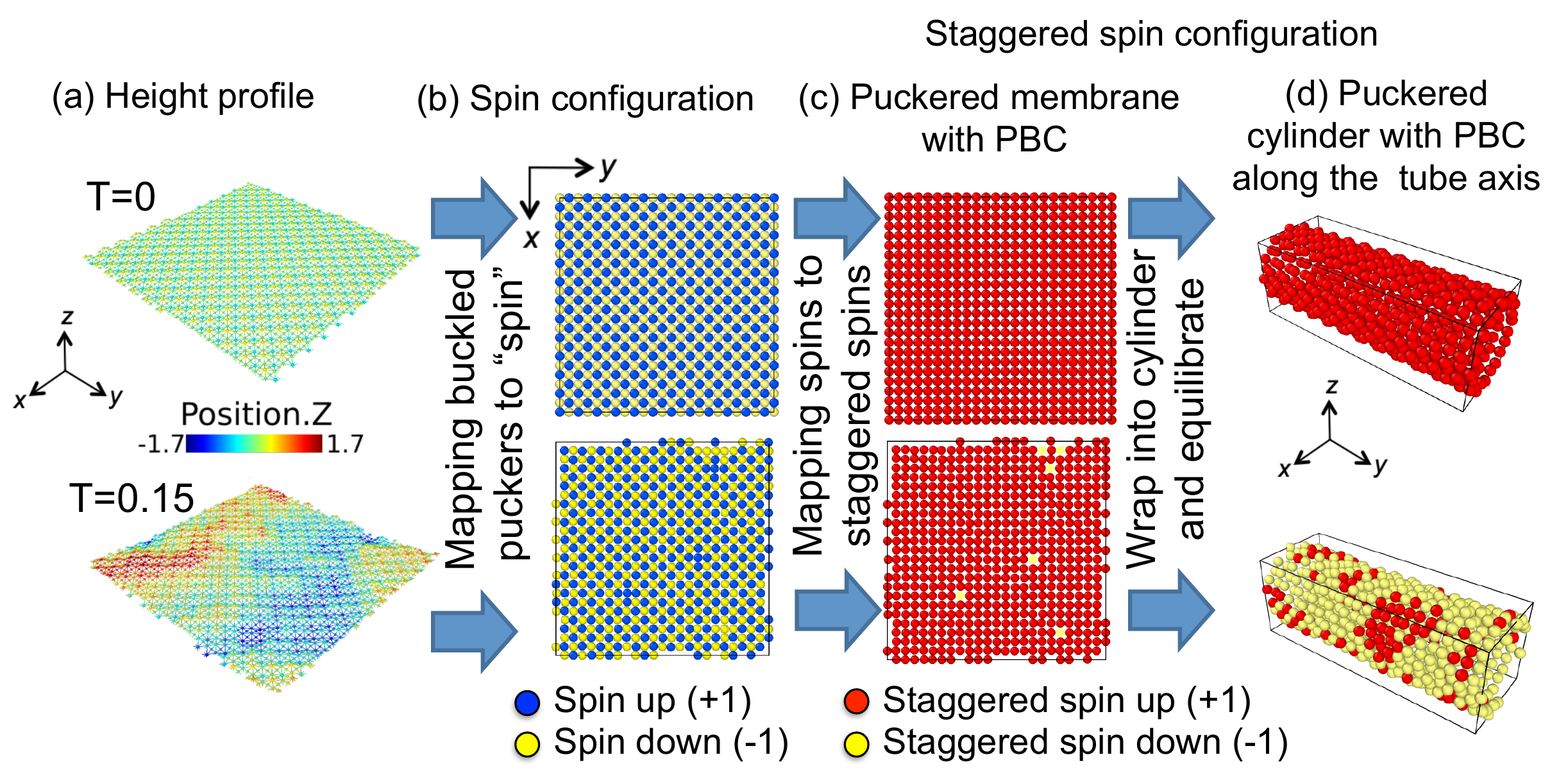}
\caption{Mapping from buckled dilation nodes to spin or staggered spin in planar and cylindrical geometries at $T=0$ (first row) and $T=0.15<T_c$ (second row), where $T_c$ is the critical temperature of the staggered (``antiferromagnetic") pucker phase transition in the planar geometry. Temperatures are measured in units of the microscopic bending rigidity $\hat{\kappa}$. (a) Height profile at equilibrium of a 48$a_0\times$48$a_0$ puckered sheet with 24$\times$24 dilation nodes with periodic boundary conditions in the $x$ and $y$ directions. The colors represent nodes' positions relative to the zero plane in units of the lattice spacing $a_0$. (b) Spin configurations associated with column (a), where dilations that buckle above the local plane formed by their neighbors are designated spin up ($\sigma=+1$, blue) and dilations that buckle below are designated spin down ($\sigma=-1$, yellow).  (c) Staggered spin configurations associated with the spins in column (b), measuring each spin's adherence to a checkerboard ordered phase. This transformation amounts to multiplying the spins on every other lattice site by $-1$  (d) The same puckered sheet as in columns (a)-(c), but now wrapped into a cylinder with periodic boundaries in the axial direction. Staggered spins are shown following energy minimization at $T=0$ or equilibration at $T=0.15$. Node positions are visualized using OVITO software~\cite{ovito}.}
\label{fig:mapping}
\end{figure*}

The total energy of our lattice model is given by
\begin{equation}
  E=\frac{k}{2}\sum_{\langle i,j\rangle}(|\pmb{r}_{i}-\pmb{r}_j|-a_{ij})^2+\hat{\kappa}\sum_{\langle \alpha, \beta \rangle}(1-\pmb{n}_{\alpha}\cdot\pmb{n}_{\beta}). 
  \label{eq:modelenergy}
\end{equation}
The first sum is over neighboring nodes and gives the stretching energy in terms of the spring constant $k$ and the rest length of the spring connecting nodes $i$ and $j$, $a_{ij}$. The second sum is over neighboring plaquettes and gives the bending energy in terms of the microscopic bending rigidity $\hat{\kappa}$. The rest lengths $a_{ij}$ are chosen to model a dilation array with each dilation separated by two lattice spacings (Fig. \ref{fig:schematics}e). The short bonds with projections lying in either the $x$ or $y$ direction have rest length $a_0$ if they are not connected to a dilation node and rest length $a_0(1+\epsilon), \epsilon>0$, if they are. The rest lengths of diagonal bonds are set so as to allow for a state with zero stretching energy in the inextensible limit. See refs. \cite{plummer2020buckling, hanakata2022anomalous} and Appendix \ref{sec:realspace} for details. 

We first simulate square sheets of area $L\times L$ (Fig. \ref{fig:schematics}f), and then compare with results for cylinders with axial lengths $L$ tuned to match their circumference, $L=2\pi R$ (Fig. \ref{fig:schematics}g). Periodic boundary conditions in the $x$ and $y$ directions are used for the planar membranes. We form cylinders by wrapping the square membranes around the $y$ axis, with periodic boundary conditions along the tube axis such that the planar membranes and the cylinders are topologically equivalent with, however, very different extrinsic curvatures. Energies are measured in units of $\hat{\kappa}$. We use $\epsilon=0.1$, $k=100\hat{\kappa}/a_0^2$ and $\hat{\kappa}=1$, which corresponds to a continuum two-dimensional (2D) Young's modulus $Y=4k/3$, a continuum bending rigidity $\kappa=\hat{\kappa}$, and a dilation F\"oppl--von K\'arm\'an number~\cite{plummer2020buckling, hanakata2022anomalous} $\gamma=\frac{16 k a_0^2\epsilon}{3\hat{\kappa}}\approx 53.3$. The elastic parameters are chosen so that buckling either up or down out of the local tangent plane is energetically preferred by dilations ($\gamma>\gamma_c\simeq 21$~\cite{hanakata2022anomalous}, Appendix \ref{sec:realspace}).

We now introduce our main simulation results graphically via Fig. \ref{fig:mapping}, though we postpone detailed discussions to Sec. \ref{sec:simulations}. In the top row, we show relaxed configurations at zero temperature obtained by minimizing energy and stress with the fast inertial relaxation engine (FIRE) algorithm~\cite{fire-algo}. Dilations embedded in a host lattice buckle in the local $z$ direction in a checkerboard pattern (Fig.  \ref{fig:mapping}a). These buckled nodes can be mapped to up and down spins (Fig. \ref{fig:mapping}b), which can then be used to determine the staggered spin (Fig. \ref{fig:mapping}c). The staggered spin variable is obtained by multiplying the spin by $(-1)^{x_i+y_i}$, where the integers $x_i$ and $y_i$ index the spin's position on the lattice. Thus the spins on one sublattice of the bipartite square lattice are multiplied by $-1$, while the others remain the same. This transformation means that if one superimposes a checkerboard on the spin configurations shown in Fig. \ref{fig:mapping}b, a spin is assigned to be staggered spin $+1$ if it is consistent with that particular checkerboard and $-1$ if it is not. Thus, both a pure staggered spin up state and a pure staggered spin down state correspond to perfect checkerboard order, with their corresponding spin configurations differing by an overall factor of $-1$. The average of the staggered spin is the staggered magnetization, the order parameter identifying the checkerboard phase. In Fig. \ref{fig:mapping}d, the surface is shown wrapped into a cylinder in real space with its staggered spin configuration superimposed. 

In the bottom row of Fig. \ref{fig:mapping}, we carry out the same set of transformations at a temperature greater than zero but less than the critical temperature $(T_c\simeq0.20)$~\cite{hanakata2022anomalous} of the staggered magnetization phase transition in the planar geometry. In Fig. \ref{fig:mapping}a, we now observe long wavelength thermal fluctuations generating out-of-plane displacements significantly greater than the dilation buckling amplitude ($\approx 0.4 a_0$). To assign spin configurations, we use the nodes' positions relative to the local planes formed by their neighbors. In Figs. \ref{fig:mapping}b and c, we observe that checkerboard order is largely maintained at $T=0.15$. However, checkerboard order is broken up for the same system when equilibrated in a cylindrical geometry! The curvature has decreased the effective critical temperature of the phase transition, and $T=0.15$ now lies in the $\mst=0$ phase. This outcome is reminiscent of the effect of a uniform external field in Ising antiferromagnets (see Appendix \ref{sec:mft} and refs.~\cite{muller1977interface, binder1980phase}). However, as we will see, the strength of this effective field is \emph{size dependent}, varying inversely as the radius of the cylinder. 

\section{Discrete real space theory}\label{sec:discrete}
To better understand these results, we now derive a microscopic field-like interaction between the curvature and the buckled dilations at zero temperature by working directly with an approximate form of the energy functional used in simulations, Eq. \ref{eq:modelenergy}. 

If we assume as a first approximation that there are no displacements tangent to the surface defined by the host lattice and only dilation nodes have displacements normal to the surface (consistent with a large F\"oppl--von K\'arm\'an number), we can express the energies of the pairs of buckled dilations in Fig. \ref{fig:schematics} solely in terms of the out-of-plane displacement, the lattice constant, and the energetic parameters. For the dense arrays studied here and in ref. \cite{hanakata2022anomalous}, applying this approximation to planar arrays leads to quantitatively accurate predictions for the buckling threshold, the height of buckled dilations, and macroscopic expansion if the lattice constant is allowed to vary (Appendices \ref{nobreathing} and \ref{breathing})~\footnote{Although these approximations are suitable for understanding the puckered arrays of Fig.~\ref{fig:mapping}, they cannot reproduce even the qualitative behavior of systems with dilations separated by more than two lattice constants---the strongly antiferromagnetic, less dense dilation arrays of ref. \cite{plummer2020buckling} would appear noninteracting under this set of assumptions, for example.}

With these assumptions, all the terms in the energy of the ferromagnetic and antiferromagnetic spin pairs on a curved host lattice (Fig. \ref{fig:schematics}c and d) are identical except for the bending energy generated by the two sets of adjacent plaquettes with normals labeled $\*n_{\alpha}$ and $\*n_{\beta}$. The contribution to the bending energy from these sets of plaquettes along a cross section line of a cylinder can be found by directly calculating the interactions between the normals (Appendix \ref{ap:curv}), which gives
\begin{equation}
\small
\begin{aligned}
2\hat{\kappa}(1-\*n_{\alpha} \cdot \*n_{\beta})&=2\hat{\kappa}\left(1+\frac{1}{\sqrt{a_0^2+f_1^2}\sqrt{a_0^2+f_2^2}} \left(f_1f_2\left(1-\frac{2 a_0^2}{R^2}\right)\right.\right.  \\ 
&\left.\left.- \frac{2a_0^2}{R}(f_1+f_2) \sqrt{1-\frac{a_0^2}{R^2}} +a_0^2 \left( \frac{2a_0^2}{R^2}-1\right)\right)\right),
\label{ebend_f_main}
\end{aligned}
\end{equation}
where $f_1$ and $f_2$ are the perpendicular displacements of the left and right dilations relative to the host lattice in Fig. \ref{fig:schematics}c,d respectively.

\subsection{Effective external field}
To see that curvature enters Eq. \ref{ebend_f_main} as an effective external field acting on an antiferromagnetic Ising model, we set $f_1=\sigma_1 f$ and $f_2=\sigma_2 f$, where $\sigma_{1,2}=\pm 1$. While a good assumption in a planar geometry, this approximation is less accurate for the cylinder---curvature breaks the up-down symmetry of the system, and $|f_1| \neq |f_2|$ in our simulations of antiferromagnetically buckled nodes. We relax this assumption in Appendix \ref{sec:discrete_expand}, and also consider the total energy of a small patch, rather than just the bending between an isolated pucker pair, with only minor changes in the results. 

Upon expanding Eq. \ref{ebend_f_main} in the limit $a_0/R \ll 1$ (small curvature or large radius) and simplifying, we find
\begin{align}
2\hat{\kappa}(1-&\*n_{\alpha} \cdot \*n_{\beta}) \approx \frac{2\hat{\kappa}}{f^2+a_0^2}\left(f^2+\frac{2a_0^4}{R^2}\right)\nonumber \\&+\frac{2 \hat{\kappa}f^2 }{f^2+a_0^2}\left(1- \frac{2 a_0^2}{R^2} \right) \sigma_1\sigma_2- \frac{4 \hat{\kappa}a_0^2 f}{R(f^2+a_0^2)}(\sigma_1+\sigma_2)\nonumber \\&+\mathcal{O}(a_0^3/R^3).
\end{align} 
After neglecting the $\sigma_j$-independent term, this contribution to the bending energy has the form of the Ising Hamiltonian $\Delta H_{12}$ (per nearest-neighbor pair of dilations) for a spin system in an external field, where $\Delta H_{12}=J_{\text{eff}}\sigma_1\sigma_2-\frac{h_{\text{eff}}}{2}(\sigma_1+\sigma_2)$, with 
\begin{align}
J_{\text{eff}}&=\frac{2 \hat{\kappa}f^2 }{f^2+a_0^2}\left(1- \frac{2 a_0^2}{R^2} \right), \\
h_{\text{eff}}&\approx\frac{8 \hat{\kappa}a_0^2 f}{R(f^2+a_0^2)}.\label{eq:field}
\end{align}
Note that the effective uniform field $h_{\text{eff}}$ is \emph{size dependent}, vanishing like the reciprocal of the cylinder radius. This calculation describes the interaction between two puckers connected along the azimuthal direction of the cylinder, as in Fig. \ref{fig:schematics}. The dilations connected along the axial direction will have an interaction of strength $J_{\rm eff}(R \to \infty)$ (Appendix \ref{sec:realspace}). We only simulate cylinders with a circumference larger than or equal to $12a_0$, for which this estimate of $J_{\text{eff}}$ is always positive, as expected for an antiferromagnet. Curvature biases the system towards positive $\sigma_1,\sigma_2$ (outward buckling), at linear order in $a_0/R$, and reduces the strength of the antiferromagnetic interaction at quadratic order in $a_0/R$. 


\subsection{Estimate of the threshold radius}\label{sec:critradius}
As the cylinder becomes more strongly curved ($1/R$ increases), the effective external field will bias the dilations to buckle away from the center of curvature, as pictured in Fig. \ref{fig:schematics}c, and the new term in the effective coupling will weaken the antiferromagnetic interaction. At some threshold radius (which will be a function of the elastic parameters), ferromagnetism will become the preferred ground state at zero temperature. A rough estimate of this threshold radius follows if we assume that the buckling magnitude $f$ is the same for both ferromagnetic and antiferromagnetic patterns. We calculate the energy difference between two small patches of antiferromagnetically and ferromagnetically puckered dilations curved into cylindrical caps (explicitly given in Eqs. \ref{eq:stretchpatch} and \ref{eq:bendpatch}), and find the energy difference per pucker,
\begin{equation}
\small
\frac{E_{\rm AFM}-E_{\rm FM}}{N_p}=- \frac{8 \hat{\kappa}f^2}{f^2+a_0^2} + \frac{8 \hat{\kappa} a_0^2 f^2 }{(a_0^2 +f^2) R^2}+\frac{8 \hat{\kappa} a_0^2 f \sqrt{1- \frac{a_0^2}{R^2}}}{(a_0^2+f^2)R}.
\label{eq:diff_AFM_FM}
\end{equation}
Upon solving Eq.~\ref{eq:diff_AFM_FM} for when $E_{\rm AFM}$ and $E_{\rm FM}$ are equal, we find a threshold radius below which outward ferromagnetic puckering dominates,  
\begin{equation}
R_{t}=\frac{a_0}{f} \left(\sqrt{f^2+a_0^2} \right),
\end{equation}
where $f$ is the local pucker amplitude. We estimate $R_{t}$ for the parameters used in simulations by substituting $f=0.374a_0$, the buckling amplitude for systems with planar periodic boundaries (Appendix \ref{breathing}, \footnote{In simulations with a cylindrical geometry, the amplitudes of antiferromagnetic and ferromagnetic buckling vary as a function of cylinder radius. The amplitude of antiferromagnetic buckling is close to the planar buckling amplitude, while the amplitude of ferromagnetic buckling can be $\sim 40\%$ lower.}), which gives 
\begin{equation}
R_{t}\approx 2.85 a_0.
\end{equation}

Although this estimate depends sensitively on our assumptions about the value of $f$, it does reveal the existence of a threshold radius in the discrete theory. In Sec. \ref{sec:simulations}, we measure the threshold radius in simulations and find a larger value, $R_{t}\approx 4 a_0$. 

\section{Nonlinear continuum elastic theory from shallow shell theory}\label{sec:continuum}
We now introduce a complementary continuum theoretical model for puckers on a cylindrical host lattice using shallow shell theory \cite{koiter2009wt, paulose, kosmrljsphere, komura1992fluctuations}. We use this model to calculate the energy in terms of the amplitudes of the staggered magnetization and magnetization buckling modes, which reveals a field-like coupling between curvature and magnetization. In contrast to the discrete model presented in Sec. \ref{sec:discrete}, which only accurately describes interactions between spins whose associated plaquettes share an edge, the complementary continuum model is most accurate in the limit of dilute dilation arrays, for which dilations are far apart and can be reasonably modeled as $\delta$-function perturbations in the preferred metric \cite{radzihovsky-PRA-44-3525-1991}. Though we only work at zero temperature in what follows, we comment on how this calculation could be extended to nonzero temperatures as well.

\subsection{Energy functional}
Consider a patch of puckered dilations with a cylindrically curved host lattice, as in Fig. \ref{fig:schematics}(c,d). For a shallow, nearly flat cylinder of radius $R$, we can parametrize the curved background surface of the cylinder, $\*r_0$, using the Monge representation, placing the origin at the top of the cylinder,
\begin{equation}
    \*r_0(x_1,x_2)=\left(x_1, x_2, Z(x_1)\right),
\end{equation}
where 
\begin{equation}
    Z(x_1)=R \left(\sqrt{1-\frac{x_1^2}{R^2}} -1 \right).
\end{equation}

Shallow shell theory assumes that the slope of the surface is small, which for our case requires
\begin{equation}
    \left| \frac{\partial Z}{\partial x_1} \right| =\left| \frac{x_1}{\sqrt{R^2-x_1^2}} \right| \ll 1,
\end{equation}
thus restricting our attention to the region close to the origin where $x_1^2 \ll R^2/2$. 
Deformations relative to the cylindrical background surface $\*r_0$ can now be decomposed into displacements tangent to the surface (in the $\hat{\*t}_1^0$ and $\hat{\*t}_2^0$ directions) and normal to the surface (in the $\hat{\*n}^0$ direction) such that
\begin{equation}
    \*r(x_1,x_2)=\*r_0+u_1 \hat{\*t}_1^0+u_2 \hat{\*t}_2^0 + f \hat{\*n}^0.
    \label{eq:config}
\end{equation}
A deformation with positive $f$ corresponds to a ``spin up" pucker with a increased radial displacement \textit{relative} to the cylindrical background surface \footnote{This is in contrast to an alternative sign convention, in which the origin is placed at the south pole of a sphere \cite{paulose, kosmrljsphere}}.

Upon applying the small slope approximation such that $\partial Z/\partial x_1 \approx -x_1/R$ and neglecting  $(\partial Z/\partial x_1 )^2$  and $f u_1 \partial Z/\partial x_1$ terms, Eq. \ref{eq:config} can be reexpressed as
\begin{equation}
    \*r(x_1,x_2)=\left( x_1+ u_1 + f\frac{x_1}{R}, x_2+u_2, Z(x_1)+f \right).\label{eq:deformed}
\end{equation}

Consistent with these approximations, an array of dilations can be modeled as a sum over $\delta$-functions at regularly spaced positions $\{\*r_i\}$ in the preferred metric tensor $g_{\alpha \beta}^0$,   \cite{radzihovsky-PRA-44-3525-1991, plummer2020buckling},
\begin{equation}
    g_{\alpha \beta}^0= \delta_{\alpha \beta} \left(1+\Omega_0 \sum_{i} \delta^2(\*r-\*r_i) \right)\equiv \delta_{\alpha \beta} \left(1+\Omega_0 c(\*r) \right),
    \label{eq:metric}
\end{equation}
where $\alpha, \beta \in \{1,2\}$, $\Omega_0$ is the extra area provided by each dilation, and $c(\*r)$ is the concentration of dilations. 

The metric tensor of the deformed or actual configuration can be found by computing $g_{\alpha \beta}=\frac{\partial \*r}{\partial x_{\alpha}}\cdot \frac{\partial \*r}{\partial x_{\beta}}$ using Eq. \ref{eq:deformed}. Thus, the strain tensor that penalizes deviations from the metric of Eq. \ref{eq:metric} is given by
\begin{align}
    &\tilde{u}_{\alpha \beta}\equiv \frac{1}{2} \left(g_{\alpha \beta} - g_{\alpha \beta}^0\right) \nonumber \\ 
    &= \frac{1}{2}\left(\frac{\partial u_{\alpha}}{\partial x_{\beta}}+\frac{\partial u_{\beta}}{\partial x_{\alpha}}+\frac{\partial f}{\partial x_{\alpha}} \frac{\partial f}{\partial x_{\beta}}\right)+\frac{f}{R} \delta_{1 \alpha} \delta_{1 \beta}-\frac{1}{2} \Omega_{0}c(\*r), 
    \label{eq:strain}
\end{align}
which defines our stretching energy in terms of the Lam\'{e} parameters,
\begin{equation}
    E_s= \frac{1}{2} \int d^2 r \left[ 2 \mu \tilde{u}_{\alpha \beta}^2 + \lambda \tilde{u}_{\gamma \gamma}^2 \right].
    \label{eq:estretch}
\end{equation}

We also impose a bending energy via the bending rigidity, $\kappa$, penalizing the square of the mean curvature \cite{seung-PRA-38-1005-1988},
\begin{equation}
\small
    E_b= \frac{\kappa}{2} \int d^2 r \left(\nabla^2 (Z(x_1)+f)\right)^2=\frac{\kappa}{2} \int d^2 r \left(\nabla^2 f - \frac{1}{R}\right)^2.
    \label{eq:ebend}
\end{equation}

The total energy is the sum of these two terms, and is in general a function of both tangential and normal displacements. However, for the purposes of our study, we are only interested in normal displacements, which determine the spin configuration. As described in Appendix \ref{sec:derive}, we can eliminate the tangential displacements at either zero or low finite temperature: At zero temperature, we minimize the energy functional with respect to tangential displacements $u_\alpha$ for a fixed function of $f$ \cite{plummer2020buckling}, and at finite temperatures, we integrate over the tangential phonon degrees of freedom in the partition function \cite{nelsonbook}. For either case, we find a relatively simple (free) energy for phonon displacements normal to the host lattice surface,
\begin{align}
    E=&\frac{\kappa}{2} \int d^{2} r\left(\nabla^{2} f-\frac{1}{R}\right)^{2}\nonumber \\ 
    &+\frac{Y}{2} \int^{\prime} d^{2} r\left(\frac{1}{2} P_{\alpha \beta}^{T} \partial_{\alpha} f \partial_{\beta} f-\frac{\Omega_{0}}{2} c(\mathbf{r})+P_{11}^{T} \frac{f}{R}\right)^{2},\label{eq:en_real}
\end{align}
where $Y=\frac{4 \mu(\mu+\lambda)}{(2\mu+\lambda)}$ is the 2D Young's modulus and $P_{\alpha \beta}^T$ is the transverse projection operator \cite{nelsonbook}. The  prime on the second integral signals that the $\*q=0$ mode is excluded.

To probe the structure of Eq. \ref{eq:en_real}, we Fourier transform the energy $E$ by introducing $f(\*q)=\frac{1}{A} \int d^2 r f(\*r)e^{-i \*q \cdot \*r}$, where $A$ is area spanned by $x_1$ and $x_2$. The Fourier transform of the dilation concentration is $c(\*q)= \frac{1}{v} \sum_{\*G} \delta_{\*q, \*G}$, in terms of $v$, the real space area of the unit cell, and a set of reciprocal lattice vectors $\{\*G\}$. Upon neglecting constants and a term of order $\left(\Omega_0/v\right)^2$, we arrive at an energy per unit area,
\begin{widetext}
\begin{align}
    \frac{E}{A}&= -\frac{Y \Omega_0}{2 v R} \sum_{\*q \neq 0 \atop \*G} P_{11}^T(\*q) f(\*q) \delta_{\*G, -\*q}+\frac{Y \Omega_{0}}{4 v} \sum_{\mathbf{q}_1+\mathbf{q}_2=\mathbf{q} \neq 0 \atop \mathbf{G}=-\mathbf{q} \neq 0} P_{\alpha \beta}^{T}\left(\mathbf{q}_1+\mathbf{q}_2\right) q_{1\alpha} q_{2\beta} f\left(\mathbf{q}_1\right) f\left(\mathbf{q}_2\right) \delta_{\mathbf{G},-\mathbf{q}_1-\mathbf{q}_2}\nonumber \\+ \frac{\kappa}{2} \sum_{\*q \neq 0}  &q^4 f(\*q) f(-\*q)
    + \frac{Y}{2 R^2} \sum_{\*q \neq 0} \left(P_{11}^T(\*q)\right)^2f(\*q)f(-\*q) -\frac{Y}{2 R}\sum_{\*q_1+\*q_2 =\*q\neq 0}P_{\alpha \beta}^T(\*q_1+\*q_2)q_{1\alpha}q_{2 \beta} f(\*q_1) f(\*q_2)P_{11}^T(\*q_1+\*q_2)f(-\*q_1-\*q_2)
    \nonumber \\
    +& \frac{Y}{8} \sum_{\mathbf{q}_{1}+\mathbf{q}_{2}=\mathbf{q} \neq 0 \atop \mathbf{q}_{3}+\mathbf{q}_{4}=-\mathbf{q} \neq 0} P_{\alpha \beta}^{T}\left(\mathbf{q}_{1}+\mathbf{q}_{2}\right) q_{1 \alpha} q_{2 \beta} f\left(\mathbf{q}_{1}\right) f\left(\mathbf{q}_{2}\right) P_{\gamma \delta}^{T}\left(\mathbf{q}_{3}+\mathbf{q}_{4}\right) q_{3 \gamma} q_{4 \delta} f\left(\mathbf{q}_{3}\right) f\left(\mathbf{q}_{4}\right).
    \label{eq:ft_f}
\end{align}
\end{widetext}
Note that linear, quadratic, cubic, and quartic terms in $f(\*q)$ are all present.

\subsection{Fourier space order parameters}\label{sec:order}
We now introduce two order parameters into our theory by associating the magnitude of the ferromagnetic buckling mode with a uniform magnetization and the magnitude of the antiferromagnetic buckling mode with a uniform staggered magnetization.

As discussed in detail in ref. \cite{plummer2020buckling}, each buckled pattern can be associated with a set of Fourier modes. For a ferromagnetic buckling pattern, the set of allowed Fourier modes is simply the reciprocal lattice vectors of the dilation superlattice:
\begin{equation}
    \*G(k_1, k_2)= \left(k_1 \frac{2\pi}{n a_0}, k_2 \frac{2 \pi}{n a_0} \right) \equiv g_0 \left(k_1, k_2 \right),\label{gimodes}
\end{equation}
where $k_1$ and $k_2$ are integers, $n a_0$ is the real space distance between dilation sites when $\Omega_0=0$, and $g_0$ is the magnitude of the smallest vector in the subspace, $2 \pi/n a_0$. Thus, the area of the real space unit cell is $v=n^2 a_0^2$. In the discrete model and simulations, $n=2$.
 
In the spirit of the nearly free electron model in solid state physics \cite{ashcroftmermin}, we approximate the ferromagnetic buckling pattern as a sum over the eight smallest nonzero reciprocal lattice vectors, $\{\*G_i\}=\{\left(\pm g_0, \pm g_0\right), \left(\pm g_0, 0\right), \left(0, \pm g_0 \right)\}$: 
\begin{equation}
     f_{\text{FM}}(\*r)=\sum_{i=1}^8 f(\*G_i) e^{i \*G_i \cdot \*r}
\end{equation}

This truncation is consistent with a square Brillouin zone that includes $|q_{x}|,|q_{y}| \leq \frac{2\pi}{n a_0}$. To determine the relevant coefficients for the Fourier modes, $\{f(\*G_i)\}$, we calculate the first eigenvector to go unstable in the limit $R \to \infty$ in the truncated basis by diagonalizing the quadratic terms in Eq. \ref{eq:ft_f}, enforcing $f^*(\*G_i)=f(-\*G_i)$, as displacements must be real \cite{plummer2020buckling}. 

We find an unstable eigenvector that has all $f(\*G_i)$ real and of the same sign. The magnitude of $f(\*G_i)$ for $|\*G_i|=g_0$ is $(1+\sqrt{5})$ times the magnitude of $f(\*G_i)$ for $|\*G_i|=g_0 \sqrt{2}$ at the buckling threshold (away from threshold the eigenvector depends on elastic parameters in the combination $\frac{Y \Omega_0}{\kappa}=\gamma$). We normalize the $f(\*G_i)$ values so that the real space peak-to-trough distance is $m a_0$, in order to match our intuitive notion of the magnetization.

Our ansatz for the real space ferromagnetic buckling deformation is thus
\begin{equation}
\begin{aligned}
   f_{\text{FM}}&(x_1,x_2) \\ &=\frac{m a_0}{4} \left( \cos(g_0 x_2)+\cos(g_0 x_1)\left(1+ \frac{2 \cos(g_0 x_2)}{1+\sqrt{5}}\right)\right) \label{eq:ffm}
\end{aligned}
\end{equation}

Similarly, the Fourier modes associated with checkerboard buckling can be found by direct calculation:
\begin{align}
    \*B(b_1, b_2)&=\left( \frac{(2 b_1+1)\pi}{n a_0},  \frac{(2 b_2+1)\pi}{n a_0} \right)\nonumber \\ 
    &\equiv \frac{g_0}{2}\left(2b_1+1, 2 b_2+1\right),\label{eq:afmfm}
\end{align}
where $b_1$ and $b_2$ are integers. 

We now approximate the antiferromagnetic buckling pattern by a sum over the four smallest nonzero wavevectors given by Eq. \ref{eq:afmfm}: $\{\*B_i\}=\{\left(\pm \frac{g_0}{2}, \pm \frac{g_0}{2} \right)\}$. The first unstable eigenvector in this basis has all $f(\*B_i)$ values equal and real. We normalize these values so that the real space peak-to-trough distance is $2 \mst a_0$. The real space antiferromagnetic buckling deformation is therefore approximated by
\begin{equation}
    f_{\text{AFM}}(\*r)=\sum_{i=1}^4 f(\*B_i) e^{i \*B_i \cdot \*r}=
    \mst a_0 \cos\left(\frac{g_0 x_1}{2}\right) \cos\left(\frac{g_0 x_2}{2}\right).
    \label{eq:fafm}
\end{equation}
We assume that, within the parameter regime studied here, other buckling modes play only a minor role in the phase behavior of the system.

\subsection{Zero temperature behavior}\label{zte}
Through this point, our calculations apply to both zero and low finite temperatures (far below the crumpling transition). If one wished to perform a low-temperature expansion, for example, one could approximate $f(\*q)$ in Eq. \ref{eq:ft_f} as a sum over wave vectors corresponding to ferromagnetic and antiferromagnetic order (Eqs. \ref{gimodes} and \ref{eq:afmfm}), wave vectors near the order parameter subspaces, generated by pucker-scale thermal fluctuations, and wave vectors with $q \ll g_0$, generated by long-wavelength thermal fluctuations. This procedure would reveal interesting temperature-dependent couplings between Fourier modes. For example, we can directly observe that the fieldlike linear term will only contribute when wave vectors with exactly the periodicity of the dilation superlattice are present, due to the $\delta$ function in that term. Similar restrictions prevent long-wavelength modes from contributing to any term proportional to $\Omega_0$, though the wave vectors close to the order parameter frequencies would certainly enter. An explicit low-temperature expansion is, however, beyond the scope of this work, and we consider only the zero-temperature behavior of the continuum theory.

At zero temperature, we assume that the deformation can be represented as a sum over the truncated subspace of ferromagnetic and antiferromagnetic modes defined in Sec. \ref{sec:order}:
\begin{equation}
    f(\*r)=m\sum_{i=1}^8 C_i e^{i \*G_i \cdot \*r}+ \mst\sum_{i=1}^4  D_i e^{i \*B_i \cdot \*r}\label{eq:modes2},
\end{equation}
where the Fourier modes $\{\*G_i\}$ and $\{\*B_i\}$ are given by Eqs. \ref{gimodes} and \ref{eq:afmfm}, and $C_i$ and $D_i$ are the constants providing the normalizations in Eqs. \ref{eq:ffm} and \ref{eq:fafm}, discussed above.

By substituting Eq. \ref{eq:modes2} into the energy per unit area, Eq. \ref{eq:ft_f}, we can examine couplings between $m$ and $\mst$. These terms would also appear as part of the low-temperature expansion procedure described above.

Upon using $n=2$ in $g_0=\frac{2 \pi}{n a_0}$ and $v=n^2 a_0^2$, we obtain a polynomial expansion in the order parameters $m$ and $\mst$,
\begin{widetext}
\begin{align}
    \frac{E(m, \mst)}{A}=&-\frac{(3+\sqrt{5}) Y \Omega_0}{128 R a_0}m+ \left( \frac{(19-\sqrt{5})Y a_0^2}{1024 R^2} + \frac{(5-\sqrt{5}) \pi^4 \kappa}{64 a_0^2} -\frac{\sqrt{5}\pi^2Y \Omega_0}{256 a_0^2}\right) m^2+\left( \frac{Y a_0^2}{32 R^2}+\frac{\pi^4 \kappa}{32 a_0^2}-\frac{ \pi^2 Y \Omega_0}{128 a_0^2}\right) \mst^2 \nonumber \\ 
    &+\frac{5 \pi^2 Y a_0}{256 R} m \mst^2+\frac{(5\sqrt{5}-5)\pi^2 Y a_0}{2048 R} m^3 +\frac{(3+\sqrt{5})\pi^4 Y}{2048}m^2 \mst^2+\frac{(7-2\sqrt{5})\pi^4 Y}{8192}m^4+\frac{\pi^4 Y}{2048} \mst^4 \label{eq:ssh_mmst}.
\end{align}
\end{widetext}
Note that we impose a cutoff on the sums over $q_1$ and $q_2$ in the quartic term such that $|q_x|, |q_y|\leq 2 \pi/n a_0$ in order to have a consistent Fourier space truncation.

Upon inspecting Eq. \ref{eq:ssh_mmst}, we see that the term $- \frac{(3+\sqrt{5})Y \Omega_0}{128 R a_0}m$ is in fact a fieldlike coupling with the same dependence on cylinder radius $R$ and sign as the effective field derived in the real space model (Eq. \ref{eq:field}). If we expand the discrete model in the amplitudes of the buckling modes (Appendix \ref{sec:discrete_expand}), we find an expansion with all of the same terms, with all the same signs as Eq. \ref{eq:ssh_mmst}, though the coefficients differ. With the exception of the $\mst^2m$ and $m^3$ terms, similar terms also appear in the usual mean-field free energy of an Ising antiferromagnet (Appendix \ref{sec:mft}).

In summary, both the continuum elastic theory and the discrete theory display couplings between the uniform magnetization and a magnetic-field-like term that scales as $1/R$. The coefficient multiplying $1/R$ in this field term differs between the two theories, as expected since the theories pertain to complementary approximations to the physics of dilation arrays. However, upon expanding the discrete model by treating the amplitudes of the buckling modes as small parameters and comparing it to the continuum model, we see that the energies of both models have the same structure. 

\section{Molecular dynamics simulations}\label{sec:simulations}
We now present simulation results for puckered cylinders at zero and finite temperature, comparing with theoretical expectations from the preceding sections when possible. Simulation details can found in Appendix~\ref{sec:md-details} and our recent work~\cite{hanakata2022anomalous}.

\subsection{$T=0$ results}
At zero temperature, we can test the prediction of the discrete model that there exists a threshold cylinder radius below which ferromagnetic order is preferred over antiferromagntic order. We simulate square membranes wrapped into cylinders with sizes in the range $12 a_0\leq L\leq 120 a_0$, or equivalently, cylinders with radii in the range $1.9a_0\leq R\leq 19.1a_0$. We initialize the pucker heights in either an antiferromagnetic or ferromagnetic configuration (with puckers pointing outward) and use the FIRE algorithm to perform structural relaxation and find the closest energy minimum \cite{fire-algo}.

\begin{figure}
\includegraphics[width=0.9\linewidth]{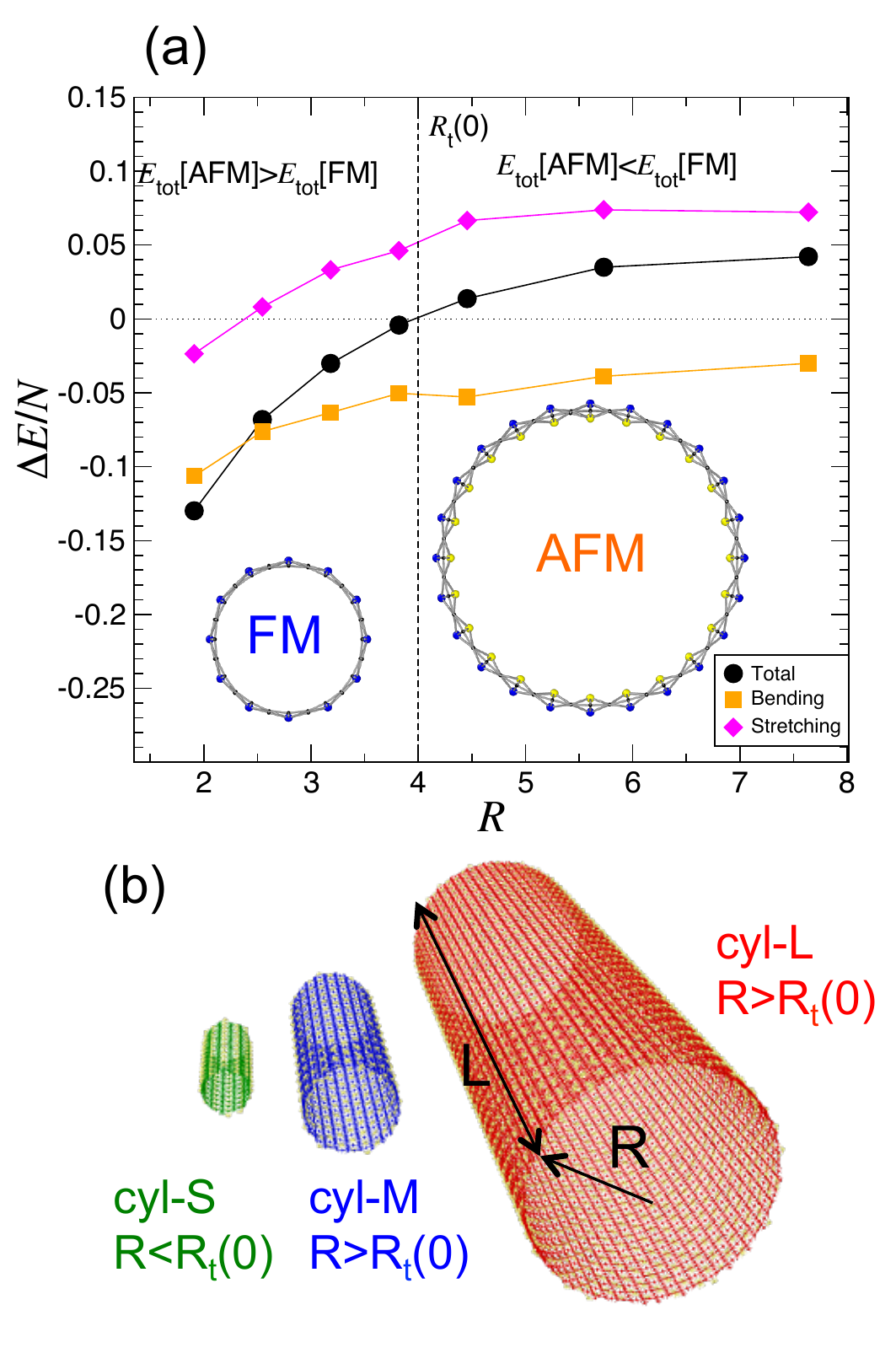}
\caption{(a) Difference in energy per site between cylinders with ferromagnetic and antiferromagnetic pucker configurations. $\Delta E$ is defined as the energy of the ferromagnetic state minus the energy of the antiferromagnetic state---therefore, for small radii/large curvatures, $E_{\text{AFM}}>E_{\text{FM}}$, $\Delta E <0 $ and ferromagnetism (with puckers pointing outward) is preferred. The dashed vertical line estimates the $T=0$ threshold radius from the simulation data, $R_t(T=0) \approx 4 a_0$. Insets show cross-sectional views of a relaxed ferromagnetic puckered cylinder of size $L=24a_0$ and a relaxed antiferromagnetic puckered cylinder of size $L=48a_0$. (b) The three representative cylinder sizes (small, medium, large) used in this work, with radii $R=3.8 a_0, 7.6 a_0$, and $19.1 a_0$ ($L= 24a_0,48a_0$, and $ 120a_0$) respectively. In the figures that follow, simulation data are colored according to the key provided by this figure: green for small cylinder results, blue for medium cylinder results, and red for large cylinder results. Periodic boundary conditions along the cylinder axis are imposed at the two ends of all cylinders.}
\label{fig:energy_difference}
\end{figure}

In Fig.~\ref{fig:energy_difference}(a) we plot the difference in total, bending, and stretching energy between ferromagnetic and antiferromagnetic states (e.g., $\Delta E_{\rm bending}=E_{\rm bending}{\rm [FM]}-E_{\rm bending}{\rm [AFM]}$), normalized by the total number of sites ($N=L^2/a_0^2$) for systems $R\lesssim 8 a_0$ ($L\leq48 a_0$).  We find the total energy of the ferromagnetic state is lower than that of the antiferromagnetic state, i.e. $\Delta E_{\rm total}=E_{\rm total}{\rm [FM]}-E_{\rm total}{\rm [AFM]}<0$, when $R\lesssim4a_0$ ($L<24 a_0$).


\subsection{$T>0$ results}
At finite temperature, we can test the prediction of both the discrete and continuum theory that the presence of curvature lowers the \emph{effective} critical temperature at which the staggered magnetization undergoes a continuous phase transition. Because our cylinders are finite, such transitions will always be rounded due to finite-size effects~\cite{cardy1996scaling, barber_1983, cardy_1988}.   


We monitor the behavior of two order parameters, introduced in Sec. \ref{sec:model} and Fig. \ref{fig:mapping}: the magnetization $m$ and the staggered magnetization $\mst$, defined as spatial averages over the up and down ``spin" configurations associated with the puckers,  
\begin{equation}
    m=\frac{1}{N_p}\sum^{N_p}_i \sigma_i,\quad \quad \mst=\frac{1}{N_p}\sum^{N_p}_i\sigma_i(-1)^{x_i+y_i},
\end{equation}
where $x_i$ and $y_i$ index the spin's lattice position, and $N_p$ is the total number of puckers.
Note that these convenient quantities differ somewhat from the buckling amplitudes we used as proxies for magnetization and staggered magnetization in Secs. \ref{sec:discrete} and \ref{sec:continuum}. Here, spins are  assigned to be either $1$ or $-1$ depending on their buckling direction, regardless of their buckling amplitude. These definitions of magnetization and staggered magnetization can be easily analyzed at finite temperature and emphasize connections with the Ising model.

\begin{figure}
\includegraphics[width=\linewidth]{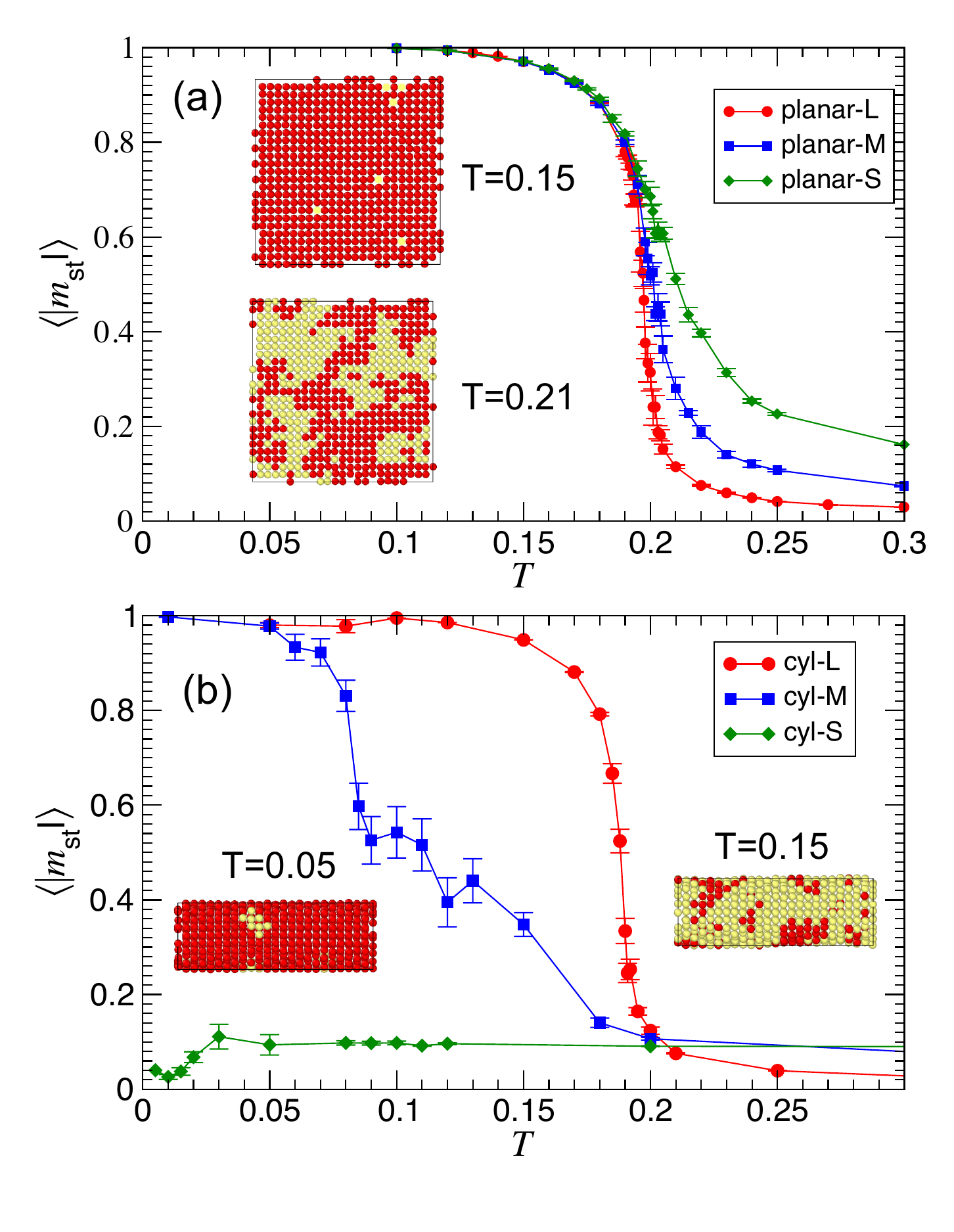}
\caption{Comparison of the average of the absolute value of staggered magnetization as a function of temperature for three different system sizes in (a) a square planar geometry with relaxed (tensionless) periodic boundary conditions and (b) a cylindrical geometry with the same dimensions. The insets to (a) compare staggered spin configurations for a $48a_0\times 48a_0$ puckered plane (24$\times$24 puckers) at $T=0.15$, below the critical temperature, and $T=0.21$, just above the critical temperature, with red and yellow circles representing up and down staggered spins, respectively. The insets to (b) compare staggered spin configurations for a $48a_0\times 48a_0$ puckered cylinder with radius $R=7.6 a_0$ at $T=0.05$ and $T=0.15$. Notice that at $T=0.15$ the puckered plane displays strong antiferromagnetic ordering whereas the puckered cylinder is already in the disordered phase, indicating that the two systems behave in a qualitatively different fashion, as if they have different critical temperatures, due to the radius-dependent ordering field from the cylindrical geometry. Error bars were calculated using the jackknife method~\cite{Young2015}}.
\label{fig:mst}
\end{figure}

In Fig. \ref{fig:mst}, we compare the staggered magnetization of three planar systems to three cylindrical systems with the same number of sites, displayed in Fig. \ref{fig:energy_difference}b. We see a number of striking differences. The planar system displays a smoothly sharpening drop in the order parameter at $T\approx 0.2$ as the system gets larger, indicative of a continuous phase transition in a finite system broadened in the usual way by conventional finite-size effects~\cite{binder-RepProgPhys-1997, sandvik-AIP-2010, barber_1983}. The large cylinder behaves similarly to the planar sheet, experiencing a smooth decay in the order parameter as a function of $T$. The medium cylinder has a more gradual decay, starting at a lower temperature, and the small cylinder displays different behavior entirely, as its ground state has zero staggered magnetization due to its high curvature. As emphasized by the insets to Fig. \ref{fig:mst}, as well as Fig. \ref{fig:mapping}, a planar system can be in the ordered phase at the same temperature that a cylinder of puckers is in the disordered phase.

\begin{figure}
\includegraphics[width=\linewidth]{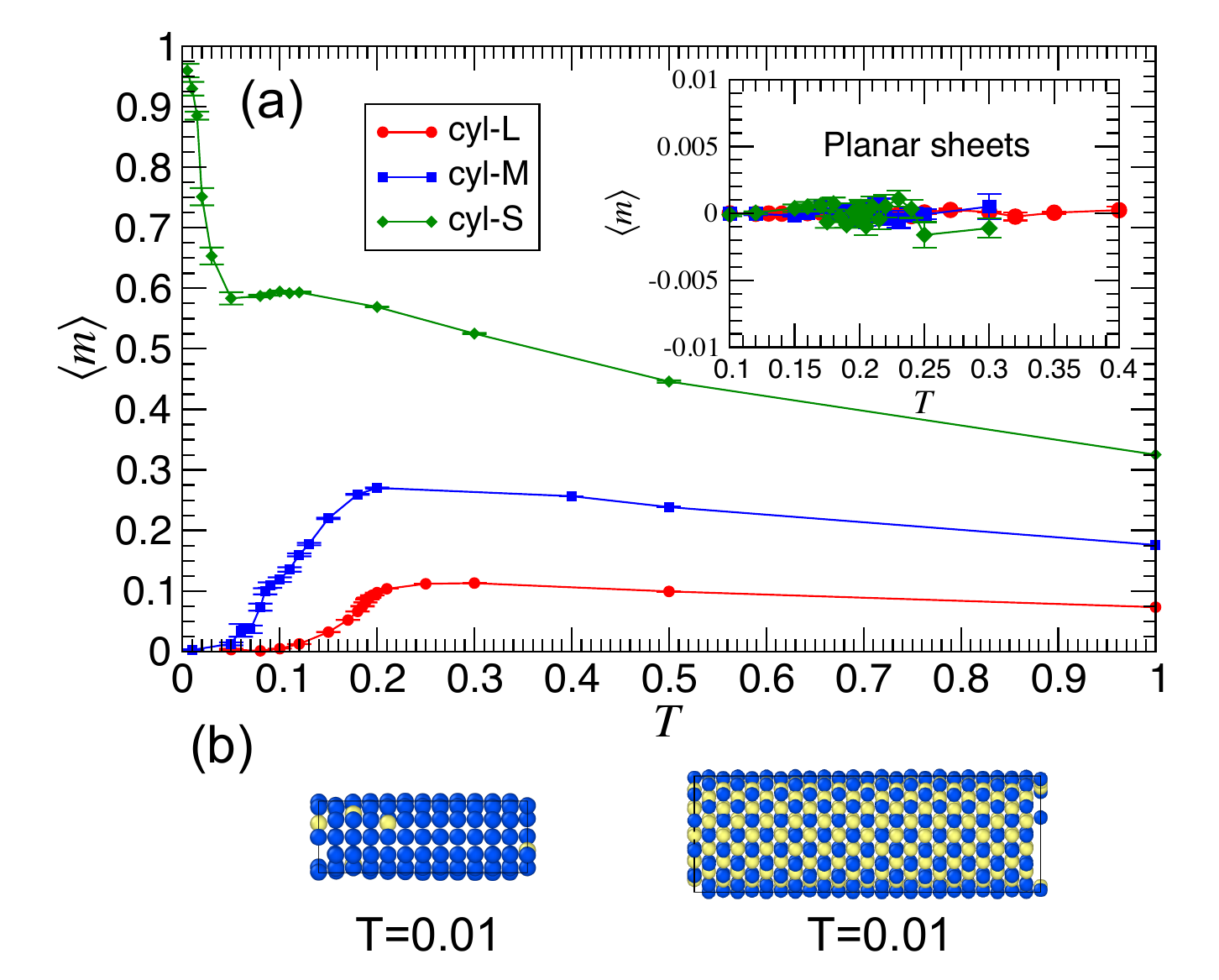}
\caption{(a) Average ferromagnetic pucker magnetization $\langle m\rangle$ of three representative cylinder sizes as a function of temperature $T$. The inset shows $\langle m\rangle\simeq0$ for puckered sheets for the three sizes we studied in a planar geometry. (b) Typical spin configurations for small and medium cylinders at $T=0.01$. The blue and yellow spheres represent spins (puckers) pointing outward and inward, respectively. At very low $T$, the small cylinder has most of its puckers pointing radially outwards ($m>0$) whereas the medium cylinder has most of its puckers pointing in and out in a checkerboard pattern ($m \approx 0$).}
\label{fig:m}
\end{figure}

Figure~\ref{fig:m} shows the average magnetization $\langle m\rangle$ of puckered cylinders and sheets as a function of $T$. For planar puckered sheets of all sizes, $\langle m\rangle\simeq0$ at any $T$. For the medium and large cylindrical systems, $\langle m\rangle$ increases from zero and reaches a small  \emph{positive} value at around $T=0.2$ before decreasing monotonically with increasing $T$. In contrast, the small cylinder has nonzero magnetization in its ground state, which decreases rapidly from $\langle m\rangle\approx 1$ for $0<T<0.1$ and continues to decrease slowly for $T>0.2$. 

Note that we plot the absolute value of $\mst$, as in our recent work~\cite{hanakata2022anomalous} and in Monte Carlo studies of Ising systems~\cite{landau-PRB-44-5081-1991}. Taking the absolute value is helpful because $\langle \mst\rangle$ averages to zero in finite-size simulations; i.e., true spontaneous symmetry breaking only occurs in the thermodynamic limit. We do not, however, take the absolute value of $m$, since the curvature of the cylinder breaks the up-down symmetry. In Fig. \ref{fig:m}, for example, $\langle m\rangle>0$ (puckers point radially outward) for all cylinders, dramatically differing from their planar counterparts. 

\begin{figure}
\includegraphics[width=\linewidth]{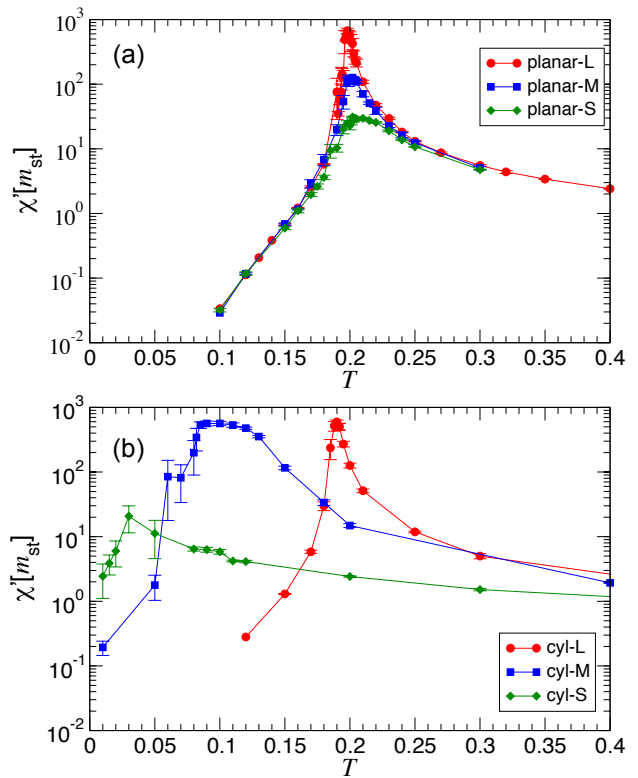}
\caption{Comparison of the thermally averaged staggered susceptibility as a function of temperature for three different system sizes in (a) a planar geometry and (b) a cylindrical geometry. The staggered susceptibility of the large cylinder is similar to that of planar systems. For the medium cylinder, however, the peak broadens and shifts to a lower temperature, with even more striking changes for the small cylinder. }
\label{fig:Xi1}
\end{figure}

Finally, we examine the susceptibility of the staggered magnetization,
\begin{equation}
\chi^\prime(\mst)= \frac{N_p}{k_{\rm B}T}\left(\langle \mst^2\rangle -\langle |\mst|\rangle^2\right),
\end{equation}
as a function of temperature for different system sizes, shown in Fig.~\ref{fig:Xi1}. In the planar systems, we again see a clear signature of critical behavior in a finite system: growing peaks in the susceptibility, with the location of the maxima converging to a well-defined  $T_c$ in the thermodynamic limit. In cylindrical systems, we observe a dramatic broadening of the peak of the susceptibility and a substantial shift in the location of the maximum as we go from a large cylinder to a medium cylinder. While the data for cylinders do not conclusively indicate the existence of a critical point obscured by finite-size effects, if we assume that this is the case we can use the energy derived in Eq. \ref{eq:ssh_mmst} to predict the shift in the critical temperature caused by the cylindrical geometry. 

Following the logic of Landau theory, we assume that the coefficients in the energy expansion given by Eq. \ref{eq:ssh_mmst} become functions of temperature once the order parameter Fourier modes are permitted to couple to thermal fluctuations, as discussed in Sec. \ref{zte}, but that no new terms appear since all terms allowed by symmetry are already present. Close to the staggered magnetization phase transition, we only consider the temperature dependence of the $\mst^2$ term. We relabel the coefficients in Eq. \ref{eq:ssh_mmst}, neglecting higher-order terms in $m$ and $\mst$, to express the free energy as
\begin{equation}
    \frac{F}{A}\approx -h(R) m + r_1(T) \mst^2 + r_2 m^2 + \zeta(R) m \mst^2 +\beta m^2 \mst^2. \label{eq:fexp}
\end{equation}
We identify the phase transition temperature within this mean-field theory as the point at which the coefficient of $\mst^2$ passes through zero, assuming that $r_1(T)=a (T-T_c)$ close to $T_c$:
\begin{equation}
    r_1(T_c(R))=a \left[T_c(R)-T_c(\infty)\right]=-\zeta(R) m -\beta m^2.\label{eq:ttc}
\end{equation}

The value of $m$ that minimizes Eq. \ref{eq:fexp} in the limit of small $\mst^2$ is $m = \frac{h(R)}{2 r_2}.$ Upon substituting this value of $m$ into Eq. \ref{eq:ttc} and the $1/R$ scalings of $h(R)$ and $\zeta(R)$ given in Eq. \ref{eq:ssh_mmst}, we find that the critical temperature for a cylinder with radius $R$ decreases as $1/R^2$.
\begin{align}
    T_c(R)&=T_c(\infty)- \frac{\zeta(R) h(R)}{2 r_2 a}-\frac{\beta h(R)^2}{4 r_2^2 a}, \nonumber \\
    &= T_c(\infty) - \text{cst.} \frac{1}{R^2}.
\end{align}

\begin{figure}
\includegraphics[width=1.1\linewidth]{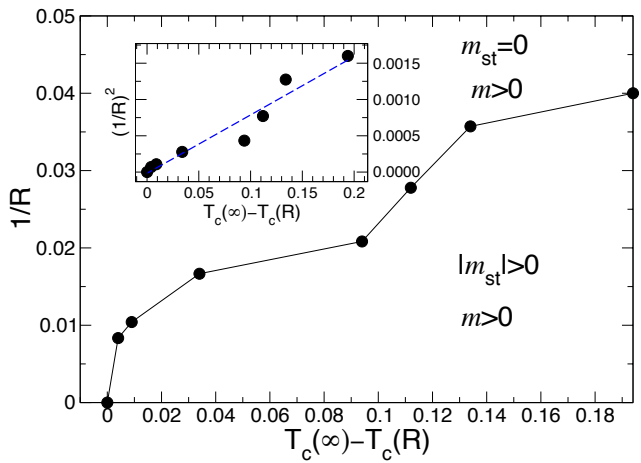}
\centering
\caption{Estimated staggered magnetization order-disorder phase boundary plotted as curvature $1/R$ vs. critical temperature of cylinders $T_c(R)$ offset by $T_c$ for $R\to \infty$. $T_c(R\to\infty)$ is the estimated $T_c$ in the thermodynamic limit of the planar system~\cite{hanakata2022anomalous}. Inset shows the linear relationship between $1/R^2$ and $T_c(\infty)$-$T_c(R)$. The dashed line is the piecewise linear fit line to the data points.}
\label{fig:phase_diagram_cyl}
\end{figure}

We test this scaling in simulations by identifying the maximum in the (possibly very broad) peak of the staggered susceptibility with $T_c(R)$. We plot the putative phase boundary obtained in this way in curvature-temperature space in Fig. \ref{fig:phase_diagram_cyl}. As shown in the inset, the shift in the critical temperature is consistent with a $1/R^2$ scaling. Since the finite size effects strengthen as the curvature increases (and the cylinder size decreases), we cannot draw firm conclusions about how quantities scale with system size without further analysis and/or simulations. We briefly discuss the effect of changing the axial length in Appendix C2.


\section{Discussion}
We have argued that the effect of curvature on arrays of buckled bistable nodes embedded in a thin elastic sheet is analogous to the effect of an external field on an Ising antiferromagnet at lowest order for large cylinder radii and to leading order in a Landau-like expansion. First, we showed that a field-like quantity scaling as $1/R$, where $R$ is the radius of curvature, couples to a ferromagnetic order parameter in two distinct theoretical models of a puckered sheet. Next, we conducted molecular dynamics simulations of puckered sheets wrapped into cylinders at zero and finite temperature and found behavior consistent with curvature acting as an external field, strongly modulated by finite-size effects. In particular, as the radius of the cylinder decreases (curvature increases), the lowest energy state switches from an antiferromagnetic configuration to a ferromagnetic configuration, and at intermediate values of the curvature we observe a shift in the effective critical temperature of the phase transition in the staggered magnetization, defined as its maximum.


In our previous work studying phase transitions in flat puckered surfaces, we were able to make precise measurements of critical exponents via finite-size scaling \cite{hanakata2022anomalous}. We did not make similar measurements in this work, since changing the size of the cylinder also changes the strength of the applied field, complicating the analysis. The correlation length in the axial direction is limited by the axial length of the cylinder, whereas the correlation length in the circumferential direction will be limited by the circumference, which couples to the effective field. We hope to investigate these subtle boundary effects in future work. 

Intriguingly, both theoretical models reveal additional terms in the energy proportional to $1/R$ that scale as $m^3$ and $\mst^2 m$, where $m$ and $\mst$ are the amplitudes of magnetization-like and staggered-magnetization-like buckling, respectively. These additional couplings to our fieldlike quantity are not present in the standard free energy expansion of an Ising model in an external field. The $m^3$ term might allow for a first-order phase transition in the magnetization to a state with negative magnetization (puckers buckled radially inwards). Although evidence of such a transition was not observed in our simulations, it would be interesting to search for by using parameters that increase the relative strength of the $m^3$ term. 

Finally, we comment on three interesting extensions of this work. First, we have focused exclusively on systems with positive dilations. Negative dilations (or contractile inclusions) in dense planar arrays have been shown to have similar phase behavior \cite{hanakata2022anomalous}, but assume profoundly different ground states in isolation \cite{xin2021instability, pezzulla2015morphing}. Both theoretical models can be generalized to negative dilations, as discussed in ref. \cite{hanakata2022anomalous} and Appendix \ref{sec:negative}. Second, cylinders allowed us to isolate the effects of background mean curvature from the more complicated (though interesting and experimentally relevant) effects of background Gaussian curvature \cite{klotz2020equilibrium, polson2021flatness}. Our shallow shell theory could be straightforwardly extended to more general curved surfaces \cite{kosmrlj-PRE-88-012136-2013, sun2021indentation}. Third, at higher temperatures, thermal fluctuations are able to crush cylindrical shells \cite{komura1992fluctuations}. The simulations presented here could be used to study whether dilation arrays can stiffen cylindrical shells and impede thermally driven collapse.

We conclude by noting that our findings are relevant for controlling the buckled phase of 2D materials such as SnO, borophane polymorphs, and many others~\cite{seixas-PRL-116-206803-2016, pacheco2019evolution, daeneke2017wafer, li2021-borophane, molle2017buckled, hanakata-PRB-96-161401-2017}. Local strains and the nature of buckling affect the electronic, optical, and spin properties of 2D materials~\cite{de2008periodically, vozmediano2010gauge,levy-Science-329-544-2010,castellanos2013local, rostami-npj2d-2-15-2018, hanakata-PRB-97-235312-2018, banerjee2020strain, makarov2021new}. Hence, the idea of using curvature as a control parameter to alter buckled structure can be applied to 2D materials on curved geometries \cite{zakharov2022shape} which can be realized experimentally in many ways, such as by rolling 2D materials into nanoscrolls~\cite{lee2006cycloaddition, goldsmith2007conductance, cui2018rolling, shenoy2010spontaneous}, adhering 2D materials onto curved substrates~\cite{de2008periodically, levy-Science-329-544-2010, castellanos2013local, banerjee2020strain}, pressurizing 2D materials with clamped boundaries~\cite{lloyd2017adhesion}, and applying in-plane strains~\cite{PhysRevApplied.18.024069}. This suggests the possibility of developing ``curvetronics," through which electronic and spin properties could be controlled via curvature.

\begin{acknowledgments}
A.P., P.Z.H., and D.R.N. acknowledge support through NSF Grant No. DMR-1608501 and via the Harvard Materials Science Research and Engineering Center, through NSF Grant No. DMR-2011754. We also thank the KITP program, ``The Physics of Elastic Films: from Biological Membranes to Extreme Mechanics," supported in part by the National Science Foundation under Grant No. NSF PHY-1748958.
HOOMD simulation input scripts and other codes are available at \url{https://github.com/phanakata/programmable-matter}. The computations in this paper were run on the FASRC Cannon cluster supported by the FAS Division of Science Research Computing Group at Harvard University.
\end{acknowledgments}

\appendix
\section{Calculations using the discrete real space model}\label{sec:realspace}

\subsection{Positive dilations with planar periodic boundaries}\label{nobreathing}
With some simplifying assumptions, we calculate the
energy of a small system of buckled positive dilations at
$T=0$, show that we can extract an effective antiferromagnetic coupling due to bending, and estimate the buckling threshold.

\begin{figure}
\begin{center}
\includegraphics[width=0.6\linewidth]{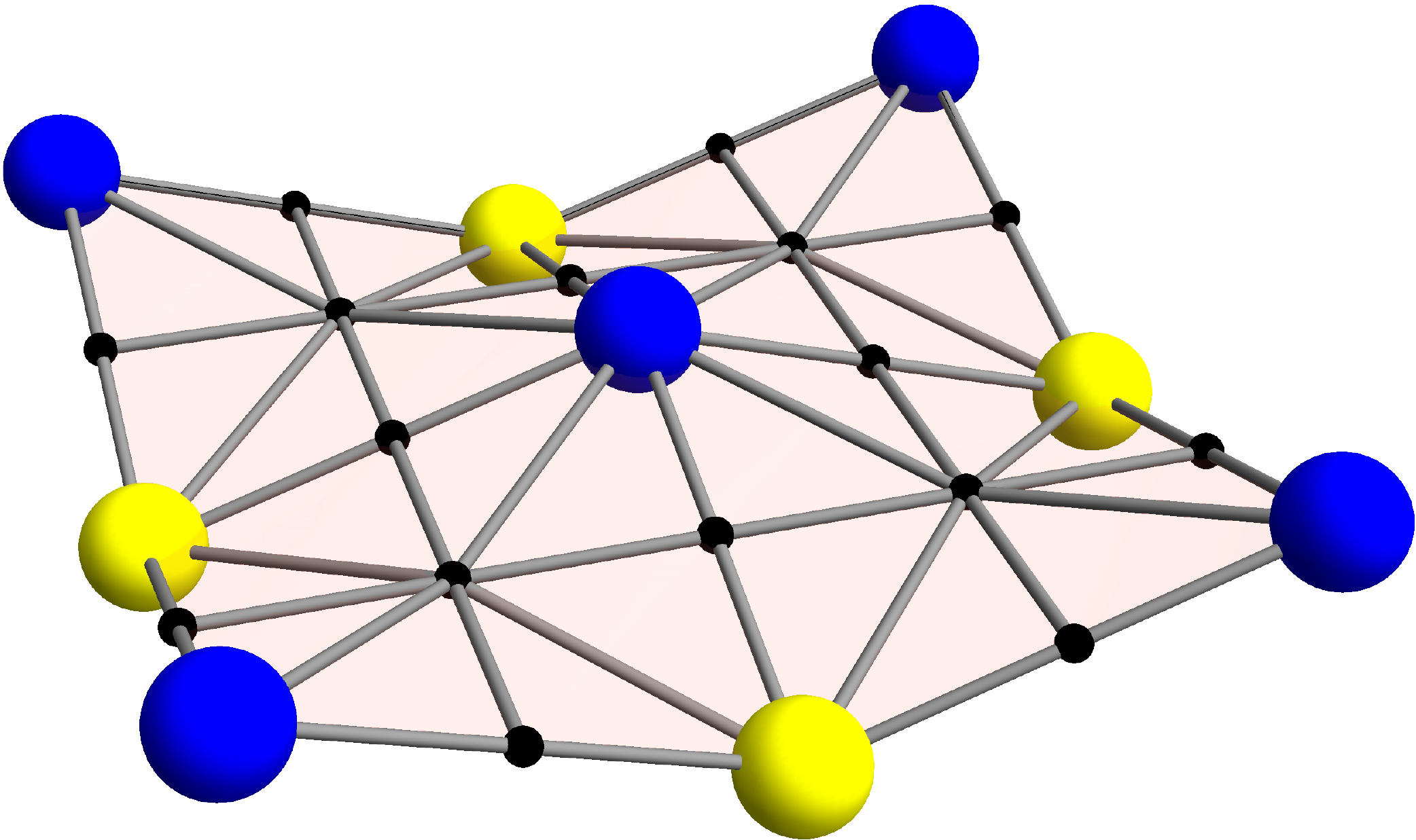}
\end{center}
\caption{{\label{fig:realspace} A perspective view of the small
    model system we consider in the checkerboard state. Blue and yellow nodes buckle in opposite
    directions. Top view is shown in Fig. \ref{fig:schematics}e. }}
\end{figure}

We consider the smallest $(0,2)$ system \cite{plummer2020buckling} for which an antiferromagnetic pattern is
allowed by the periodic boundary conditions, pictured in
Fig. \ref{fig:realspace}. Because of the boundaries, there are only
four independent dilations in the system. We make the following simplifications.

\begin{enumerate}
\item We set all in-plane displacements $u_x$ and $u_y$ to zero. 
\item We assume that the blue nodes in Fig. \ref{fig:realspace} all have height $\sigma_1 f$ and the yellow nodes have height $\sigma_2 f$, where $\sigma_{1,2}=\pm 1$ and $f>0$ is a positive height displacement. This assumption restricts us to studying either ferromagnetic or antiferromagnetic configurations (see ref. \cite{plummer2020buckling} for a discussion of other states). 
\item We require that only the dilation nodes have nonzero out-of-plane
  displacement. 
\end{enumerate}

Bending and stretching energy is now calculated using the discrete form of the energy, Eq. \ref{eq:modelenergy}. The preferred length of the bonds lying along the $x$ and $y$ directions connected to positive dilations is $a_0(1+\epsilon)$ and the
corresponding length of the diagonal bonds is
$a_0\sqrt{2+2\epsilon+\epsilon^2}$, constructed so as to allow for a
stress-free prismatic limit \cite{plummer2020buckling}. The stretching energy is
\begin{align}
E_{\text{stretch}}= &8 k \left(\sqrt{f^2+a_0^2} - a_0(1+\epsilon)\right)^2\nonumber \\ &+ 8 k \left(\sqrt{f^2+2a_0^2} - a_0\sqrt{2+2\epsilon+\epsilon^2}\right)^2,
\end{align}
where $k$ is the spring constant of the lattice model.
The stretching energy is minimized when $f=a_0\sqrt{2\epsilon+\epsilon^2}$, independent of $\sigma_1$ and $\sigma_2$. 

The bending energy is calculated by explicitly computing the normals to the triangular plaquettes in Fig. \ref{fig:realspace}, with the result
\begin{align}
E_{\text{bend}}=&16 \hat{\kappa}\left(1+\frac{f^2 \sigma_1\sigma_2-a_0^2}{f^2+a_0^2} \right) + 16 \hat{\kappa} \left(1-\frac{a_0^2}{f^2+a_0^2}\right)\nonumber \\ =&16 \hat{\kappa} \frac{f^2 (2+\sigma_1\sigma_2)}{f^2+a_0^2},
\label{ebend}
\end{align}
where the first term in the first line comes from bending across hinges formed
by short bonds, and the second term from bending across hinges
formed by long bonds. The bending energy is minimized when $f=0$ (i.e., the system is flat). 

\subsubsection{Effective antiferromagnetic coupling}
We observe that the bending energy given by Eq. \ref{ebend} has a contribution from the interaction between neighboring buckled dilations that is exactly of the form of an Ising coupling $\sigma_1\sigma_2$. The interaction term leads to an Ising Hamiltonian $\sum_{\langle i,j \rangle} J_{\text{eff}} \sigma_i \sigma_j$, where we sum over a square of four nearest-neighbor bonds connecting our puckers, with
\begin{equation}
J_{\text{eff}}= \frac{2 \hat{\kappa}}{1+a_0^2/f^2} \geq 0.
\end{equation}
$J_{\text{eff}}$ is zero when $f=0$, since there is no interaction between dilations in the flat state (bending energy is zero). When $f \neq 0$, $J_{\text{eff}}$ is strictly positive, confirming an effective antiferromagnetic interaction.

If we assume that $f \sim a_0 \sqrt{\gamma-\gamma_c}$ close to the buckling threshold $\gamma_c$ \cite{plummer2020buckling},
\begin{equation}
J_{\text{eff}} \sim \frac{\hat{\kappa}(\gamma-\gamma_c)}{\gamma-\gamma_c+1} \sim \hat{\kappa} (\gamma-\gamma_c),
\end{equation}
when $\gamma-\gamma_c \ll 1$.

\subsubsection{Buckling threshold}
The competition between bending and stretching energies decides whether the flat or buckled state is preferred, and allows an estimate of $\gamma_c$, the buckling threshold. The vanishing of the second derivative of the total energy $E(f)=E_{\text{stretch}}+E_{\text{bend}}$ with respect to $f$, evaluated at $f=0$, determines when $E(f)$ becomes a double-well potential and the flat state becomes unstable. The condition that $E(f)$ is a quartic polynomial for small $f$ is thus
\begin{equation}
k a_0^2 \left(-2 +2 \epsilon + \sqrt{4+2 \epsilon(2+\epsilon)} \right)=4(2+\sigma_1\sigma_2)\hat{\kappa} .
\end{equation}
If we neglect terms of order $\epsilon^2$ and eliminate $k$, $\hat{\kappa}$, and $\epsilon$ in favor of their macroscopic analogs $Y=\frac{4 k}{3}$, $\kappa=\hat{\kappa}$, and $\Omega_0=4 a_0^2 \epsilon$,
\begin{equation}
\gamma_c=\frac{Y |\Omega_0^c|}{\kappa}=\frac{64}{9}(2+\sigma_1\sigma_2).
\end{equation}
This threshold is first reached for antiferromagnetic buckling, $\sigma_1\sigma_2=-1$, when $\gamma_c=\frac{64}{9}\approx 7.11$. This result underestimates the threshold measured in simulations of a $(0,2)$ array, $\gamma_c=20.8$, because disallowing in-plane phonons makes the flat state artificially expensive (its ``breathing mode" is not permitted). We note that this treatment does correctly reproduce the finding that the antiferromagnetic state buckles before the ferromagnetic state as $\gamma$ is increased.

For an alternative continuum treatment of the antiferromagnetic interaction between two positive dilations in a thin elastic sheet, see ref. \cite{oshri2020buckling}.

%

\subsection{Incorporating unit cell expansion}\label{breathing}
In order to make our simplified real space model more realistic, we now allow the system to lower its energy by expanding or contracting uniformly. We thus scale the lattice constant $a_0$ by a factor $\eta$. All other assumptions of the previous section are unchanged. 

The stretching energy for the system in Fig. \ref{fig:realspace} becomes
\begin{align}
&E_{\text{stretch}}=8 k \left(\sqrt{f^2+\eta^2 a_0^2} - a_0(1+\epsilon)\right)^2\nonumber \\&+ 8 k \left(\sqrt{f^2+2\eta^2a_0^2} - a_0\sqrt{2+2\epsilon+\epsilon^2}\right)^2+8k a_0^2(\eta-1)^2. 
\label{estretch}
\end{align}
The bending energy becomes
\begin{equation}
E_{\text{bend}}=16 \hat{\kappa} \frac{f^2 (2+\sigma_1\sigma_2)}{f^2+\eta^2a_0^2}. 
\label{ebend_br}
\end{equation}

We now estimate a more accurate buckling threshold by first computing the value of $\eta$ that minimizes the energy in the flat state by solving $\frac{\partial E}{\partial \eta}\big|_{f=0}=0$. To linear order in $\epsilon$, we find
\begin{equation}
\eta\big|_{f=0}= 1+\frac{\epsilon}{2}.
\end{equation}
This result is consistent with the finding in Ref. \cite{plummer2020buckling} that $\Omega_0=4 a_0^2 \epsilon$. 

We then calculate $\frac{\partial^2 E}{\partial f^2}$ and evaluate at $f=0$ and $\eta=1+\frac{\epsilon}{2}$.
This second derivative now vanishes when
\begin{equation}
ka_0^2\left(2+\epsilon\right) \left(-2+\sqrt{4+2\epsilon (2+\epsilon) }\right) = 8 (2+\sigma_1\sigma_2) \hat{\kappa} .
\end{equation}
As above, we neglect terms of order $\epsilon^2$ and eliminate $k$, $\hat{\kappa}$, and $\epsilon$ to find two distinct puckering thresholds, one for ferromagnetism ($\sigma_1 \sigma_2=1$) and one for antiferromagnetism ($\sigma_1 \sigma_2=-1$)
\begin{equation}
\gamma_c=\frac{64}{3}(2+\sigma_1\sigma_2).
\end{equation}
The instability to antiferromagnetism ($\sigma_1 \sigma_2=-1$) again occurs first for increasing $\gamma$, with $\gamma_c=\frac{64}{3} \approx 21.3$, very close to the value measured in simulations, $\gamma_c =20.8$. 

\begin{figure}
\begin{center}
\includegraphics[width=\linewidth]{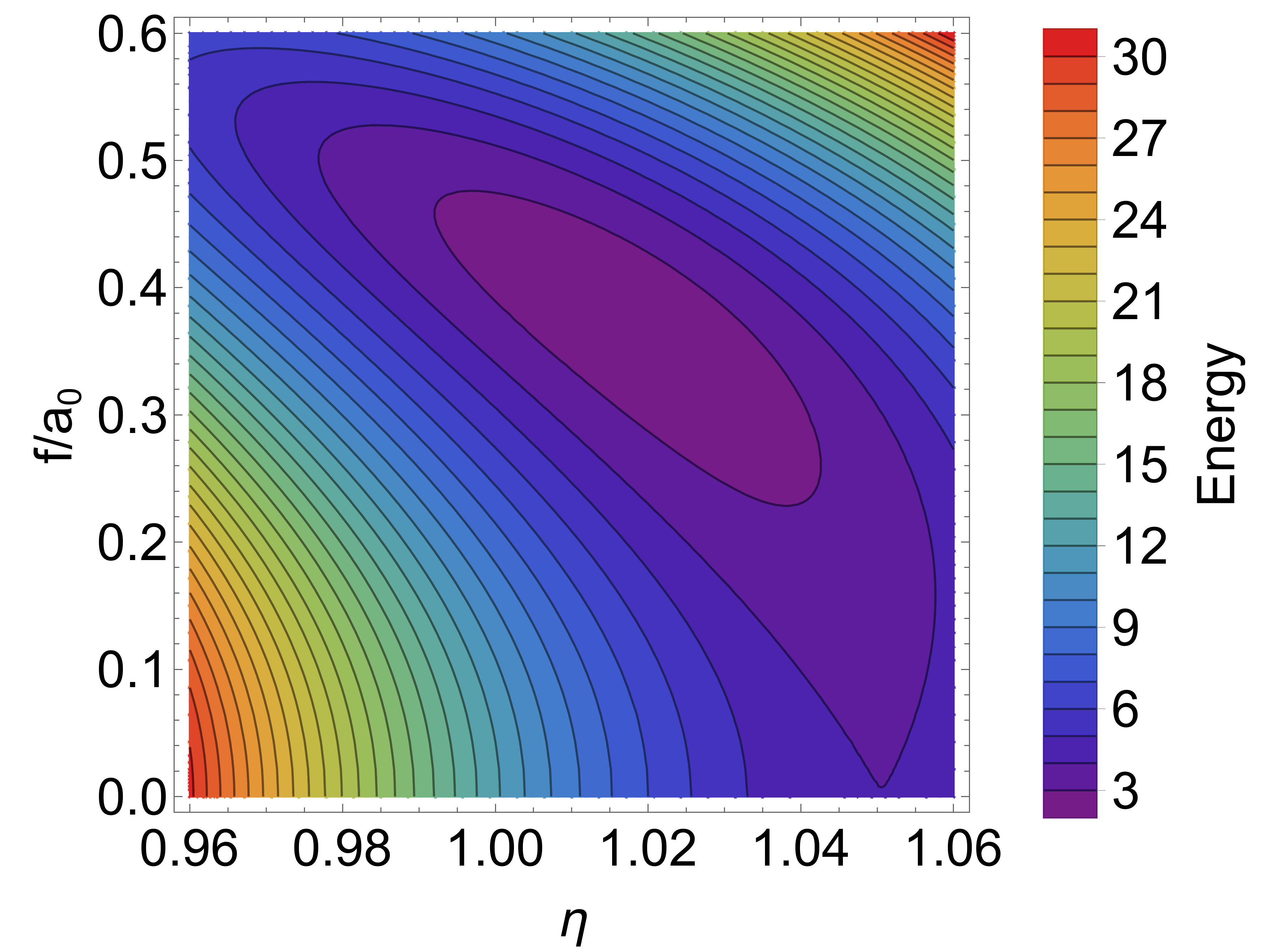}
\end{center}
\caption{{\label{fig:contour} Contour plot of the total energy of a system of four antiferromagnetic dilations as a function of the dilation factor $\eta$ and the height of the buckled dilations $f$. Energy is measured in units of $\hat{\kappa}$, the bending rigidity. $\epsilon=0.1, k=100 \hat{\kappa}/{a_0^2}$. The minimum occurs in the purple region, when $\eta=\eta_{\rm min}=1.017$ and $f=f_{\rm min}=0.374 a_0$.}}
\end{figure}

Away from the buckling threshold, we can numerically minimize the energy with respect to $\eta$ and $f$ (see Fig. \ref{fig:contour}). For the parameters used in the main text, $\epsilon=0.1, k=100 \hat{\kappa}/{a_0^2}$, we find that $\eta_{\text{min}}=1.017$ and $f_{\text{min}}=0.374 a_0$. These values are identical to those measured in simulations, where $\eta=1.017$ and $f=0.374 a_0$.

In summary, for dense, $(0,2)$ arrays, this simplified model, focused on just four nearest-neighbor puckers, can quantitatively reproduce key simulation results. This accuracy does not carry over to more dilute arrays such as those studied in Ref. \cite{plummer2020buckling}, as neighboring spins become uncoupled when their associated plaquettes do not share an edge under these assumptions.

\subsection{Effect of curvature}\label{ap:curv}
Building on the model introduced in Appendix \ref{nobreathing} (with a fixed lattice constant for simplicity) at $T=0$, we now show that a small imposed curvature leads to an effective external field term, and estimate the radius of curvature below which ferromagnetic puckers are preferred.

We start by explicitly calculating the normals $\*n_\alpha$ and $\*n_\beta$ labeled in the curved, cylindrical geometry of Fig. \ref{fig:schematics}c. Consider the three (nondilated) nodes that lie along the dotted green circle. Their positions in the $(x,z)$ plane are 
\begin{align*}
\*{r}_1&=R(-\sin(\Delta \theta), \cos(\Delta \theta)),\\
\*r_2&=R(0,1),\\
\*r_3&=R(\sin(\Delta\theta), \cos(\Delta \theta)),
\end{align*}
where $R$ is the radius of the circle that defines a cylindrical cross section. Upon assuming that the distance between $\*r_1$ and $\*r_2$ (and $\*r_2$ and $\*r_3$) is $2a_0$, we have $\Delta \theta= 2 \sin^{-1}(a_0/R)$. 

We assume that the left dilation in Fig. \ref{fig:schematics}c is displaced from the midpoint of $\*r_1$ and $\*r_2$ a distance $f_1$, and the right dilation is displaced from the midpoint of $\*r_2$ and $\*r_3$ a distance $f_2$. Their positions in the $(x,z)$ plane are, respectively,
\begin{align*}
\*p_1&=\left( -a_0 \sqrt{1-\frac{a_0^2}{R^2}}-\frac{a_0}{R}f_1, R-\frac{a_0^2}{R} +f_1\sqrt{1-\frac{a_0^2}{R^2}}\right),\\
\*p_2&=\left( a_0 \sqrt{1-\frac{a_0^2}{R^2}}+\frac{a_0}{R}f_2, R-\frac{a_0^2}{R} +f_2\sqrt{1-\frac{a_0^2}{R^2}}\right).
\end{align*}
The normal to the line formed by $\*p_1$ and $\*r_2$ is
\begin{equation}
\*n_{\alpha}=\frac{1}{\sqrt{a_0^2+f_1^2}}\left(-\frac{a_0^2}{R} +f_1\sqrt{1-\frac{a_0^2}{R^2}}, a_0 \sqrt{1-\frac{a_0^2}{R^2}}+\frac{a_0}{R}f_1\right),
\end{equation}
and the normal to the line formed by $\*p_2$ and $\*r_2$ is
\begin{equation}
\*n_{\beta}=\frac{1}{\sqrt{a_0^2+f_2^2}}\left(\frac{a_0^2}{R} -f_2\sqrt{1-\frac{a_0^2}{R^2}}, a_0 \sqrt{1-\frac{a_0^2}{R^2}}+\frac{a_0}{R}f_2\right).
\end{equation}

The two sets of adjacent plaquettes with normals $\*n_{\alpha}$ and $\*n_{\beta}$ thus contribute a term to the bending energy between neighboring plaquettes of the form
\begin{align}
2\hat{\kappa}(1-\*n_{\alpha} \cdot \*n_{\beta})&=2\hat{\kappa}\Bigg(1+\frac{1}{\sqrt{a_0^2+f_1^2}\sqrt{a_0^2+f_2^2}} \Big(f_1f_2(1-2 x^2) \nonumber \\  
&- 2 a_0(f_1+f_2)x \sqrt{1-x^2} +a_0^2 (2 x^2-1)\Big)\Bigg),
\label{ebend_f}
\end{align}
where $x=a_0/R$.
We assume, for relatively small bends $a_0/R \ll 1$, that the other terms in the bending energy and stretching energy (e.g., Eqs. \ref{ebend} and \ref{estretch}) are unchanged and remain independent of the cylinder radius $R$.

\subsubsection{Comparison with free energy expansions}\label{sec:discrete_expand}

We now demonstrate that the energy derived using the real space model has a similar structure to the energy derived using shallow shell theory by expanding the energy in terms of the amplitudes of the ferromagnetic and antiferromagnetic buckling modes. 

We consider all sources of bending and stretching energy (rather than just the single term considered above) for the small system pictured in Fig. \ref{fig:realspace} wrapped into a cylindrical cap as in Fig. \ref{fig:schematics}c,d. We assume dilations in blue have a height $f_1$ measured relative to the tilted plane formed by their neighbors (the base of the square pyramid with the dilation node at the vertex), and similarly dilations in yellow have a height $f_2$. We assume that all of the neighbor nodes of a given dilation lie in the same plane to simplify the calculation, but no longer require $|f_1|=|f_2|$. We also rescale $f_{1,2}$ by $a_0$ to make these quantities dimensionless.

The stretching energy is the same as in Appendix \ref{nobreathing}, generalized to two different pucker heights: 
\begin{align}
    E_{\text{stretch}}=&4 ka_0^2 \left(\sqrt{f_1^2+1}-(1+\epsilon)\right)^2\nonumber\\&+4 ka_0^2\left(\sqrt{f_1^2+2}-\sqrt{2+2\epsilon+\epsilon^2}\right)^2\nonumber\\&+4  ka_0^2 \left(\sqrt{f_2^2+1}-(1+\epsilon)\right)^2\nonumber\\&+4 ka_0^2\left(\sqrt{f_2^2+2}-\sqrt{2+2\epsilon+\epsilon^2}\right)^2.\label{eq:stretchpatch}
\end{align}

The bending energy has a term that is unchanged relative to the flat case corresponding to bending within a pyramid. The remaining source of bending energy is the relative rotation of neighboring plaquettes from different pyramids. Some neighboring pyramids experience additional rotation due to the underlying curvature. Our final result for the bending energy is
\begin{equation}
\small
\begin{aligned}
E_{\text{bend}}=&8\hat{\kappa}\left(\frac{f_1^2}{f_1^2+1}+\frac{f_2^2}{f_2^2+1}\right)+8 \hat{\kappa} \left(1- \frac{1 -f_1 f_2}{\sqrt{1+f_1^2}{\sqrt{1+f_2^2}}} \right)\\&+8 \hat{\kappa}\Bigg(1+\frac{1}{\sqrt{1+f_1^2}\sqrt{1+f_2^2}} \Big(f_1f_2(1-2 x^2)  \\&- 2 (f_1+f_2)x \sqrt{1-x^2} + (2 x^2-1)\Big)\Bigg),\label{eq:bendpatch}
\end{aligned}
\end{equation}

where $x=a_0/R$ as in Eq. \ref{ebend_f}.

We now expand $x$ to linear order, $\epsilon$ to linear order, and $f_1$ and $f_2$ to quartic order. These approximations are most accurate for small dilations on weakly curved surfaces ($a_0/R \ll 1$) close to the buckling threshold. Then, we define the amplitude of the ferromagnetic and antiferromagnetic buckling modes respectively as
\begin{align}
    m&=\frac{1}{2}(f_1+f_2),\label{eq:m}\\
    \mst&=\frac{1}{2}(f_1-f_2)\label{eq:mst}.
\end{align}
Upon substituting these expressions in to the expansion and dividing by $N_p v=4N_p a_0^2=A$, we find the energy density as a function of these two order parameters:
\begin{align}
        \frac{E}{A}\approx& -\frac{2 \hat{\kappa}}{R a_0} (m-\mst^2 m-m^3) +\left(\frac{3 \hat{\kappa}}{a_0^2}-\frac{3 k \epsilon}{4}\right) m^2\nonumber \\&+\left(\frac{ \hat{\kappa}}{a_0^2}-\frac{3 k \epsilon}{4}\right) \mst^2+\left(\frac{3k}{16}(6+5\epsilon) -\frac{10 \hat{\kappa}}{a_0^2}\right) \mst^2 m^2 \nonumber \\&+ \left(\frac{k}{32}(6+5\epsilon) -\frac{3 \hat{\kappa}}{a_0^2} \right) m^4+ \left(\frac{k}{32}(6+5\epsilon) -\frac{ \hat{\kappa}}{a_0^2} \right) \mst^4 \label{eq:discrete_mmst}.
\end{align}


Similar to our results for shallow shell theory, this expansion has a fieldlike term linear in $m$ that scales as $1/R$, and quadratic terms that become negative as a function of $\gamma$ (agreeing with previous results in the absence of unit cell expansion: $\gamma_c^{FM}=64/3$ and $\gamma_c^{AFM}=64/9$). Higher-order terms in $m$ and $\mst$ could be required for stability at intermediate values of $\gamma$. 

\subsection{Comments on negative dilations}\label{sec:negative}
Though a complete treatment would be beyond the scope of this paper, we now briefly explore how these real space models can be adapted to planar arrays of negative dilations, studied in ref. \cite{hanakata2022anomalous}.

\begin{figure}
\begin{center}
\includegraphics[width=\linewidth]{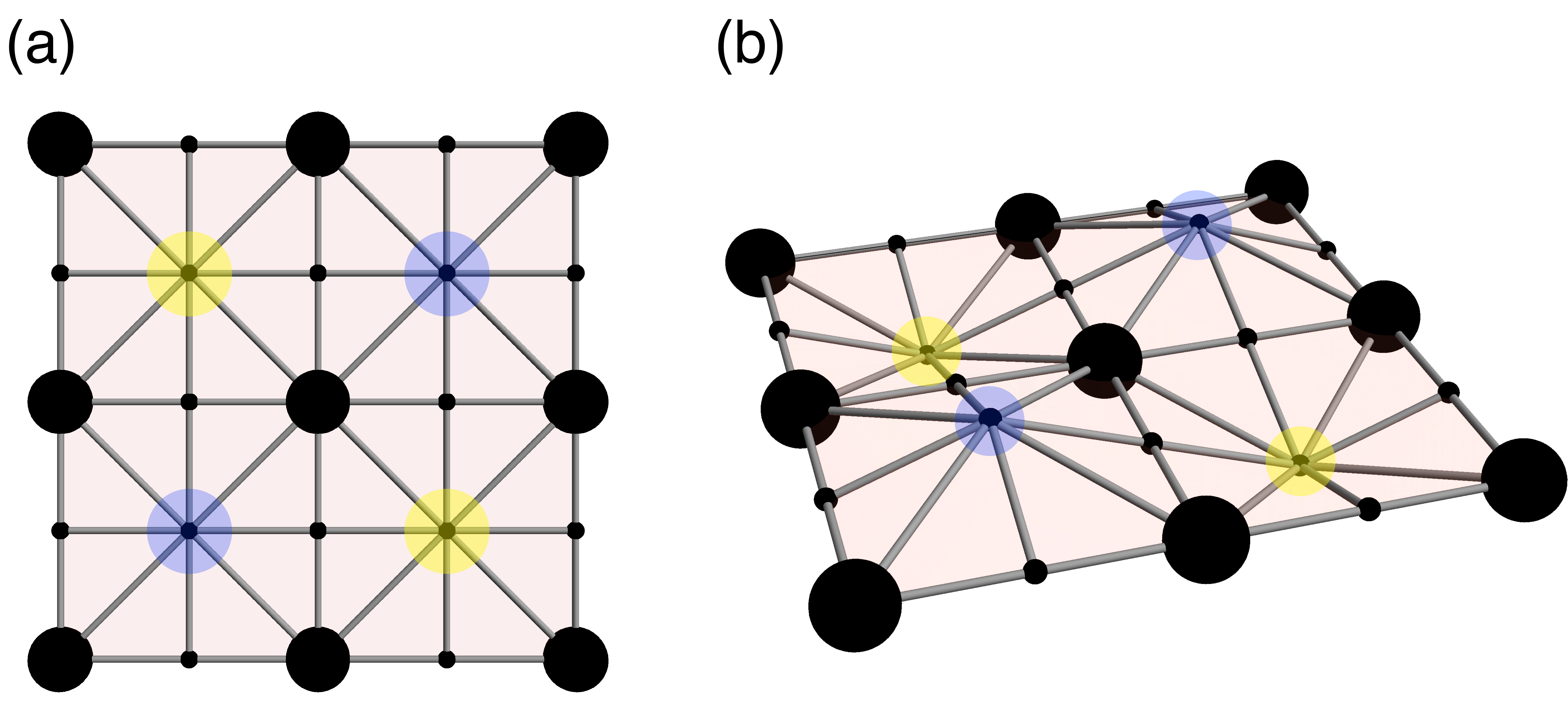}
\end{center}
\caption{{\label{fig:realspacestitch} (a) A top-down view of a small $(0,2)$ array of negative dilations (shown as large black spheres along the boundary and at the center). The nodes on the lattice dual to the dilation superlattice (highlighted in blue and yellow) buckle in a checkerboard pattern, while the dilation nodes themselves remain in plane. (b) The checkerboard state for the buckled negative dilation array viewed in perspective.
}}
\end{figure} 

When a $(0,2)$ array of negative dilations buckles, the
sites with significant out-of-plane displacement from the average height are the host lattice sites dual to the dilation superlattice (highlighted
in blue and yellow in Fig. \ref{fig:realspacestitch}), rather than the dilations themselves. We recalculate the energy of the small system with negative dilations shown in Fig. \ref{fig:realspacestitch},
assuming that the blue and yellow highlighted nodes are the only nodes with out-of-plane displacements, taking values of $\pm f$ in either a
checkerboard or ferromagnetic configuration. 

If we assume that the lattice constant is fixed, as we do in Appendix \ref{nobreathing}, we find that the flat state is always stable, even for arbitrarily large negative dilations. A global contraction seems to be a necessary condition for buckling in negative dilation arrays, in contrast to positive dilation arrays. We therefore allow for a breathing mode by multiplying the lattice constant $a_0$ by a factor $\eta$, as in Appendix \ref{breathing}, with no other in-plane displacements permitted. The stretching energy is then
\begin{equation}
\small
\begin{aligned}
&E_{\text{stretch}}=8 k \left(\sqrt{f^2+\eta^2 a_0^2} - a_0\right)^2 \\&+ 8 k \left(\sqrt{f^2+2\eta^2a_0^2} - a_0\sqrt{2+2\epsilon+\epsilon^2}\right)^2+8k a_0^2(\eta-1-\epsilon)^2. \label{eq:stitchstretch}
\end{aligned}
\end{equation}

The stretching energy differs from Eq. \ref{estretch} because the bonds connected to the displaced nodes have different rest lengths. Note that $\epsilon<0$ in Eq. \ref{eq:stitchstretch}, as is appropriate for negative dilations. We find that the bending energy is unchanged from Eq. \ref{ebend_br}
\begin{equation}
E_{\text{bend}}=16 \hat{\kappa} \frac{f^2 (2+\sigma_1\sigma_2)}{f^2+\eta^2a_0^2}.
\end{equation}

Following the same steps as in Appendix \ref{breathing}, minimizing the energy with respect to $\eta$ gives, to linear order in $\epsilon<0$,
\begin{equation}
\eta\big|_{f=0}= 1+\frac{\epsilon}{2}.
\end{equation}

We then calculate $\frac{\partial^2 E}{\partial f^2}$ and evaluate at $f=0$ and $\eta=1+\frac{\epsilon}{2}$.
This second derivative vanishes when
\begin{equation}
ka_0^2\left(2+\epsilon\right) \left(-2-2\epsilon+\sqrt{4+2\epsilon (2+\epsilon) }\right) = 8 (2+\sigma_1\sigma_2) \hat{\kappa} .
\end{equation}
As above, we neglect terms of order $\epsilon^2$, which leads to
\begin{equation}
-ka_0^2\epsilon = 8(2+\sigma_1\sigma_2) \hat{\kappa}.
\end{equation}
Upon eliminating $k$, $\hat{\kappa}$, and $\epsilon$, we find
\begin{equation}
\gamma_c=\frac{Y |\Omega_0^c|}{\kappa}=\frac{64}{3}(2+\sigma_1\sigma_2).
\end{equation}
This buckling threshold for $\epsilon<0$ thus again occurs first for antiferromagnetism ($\sigma_1 \sigma_2=-1$) and has the same magnitude as the corresponding threshold for $\epsilon>0$ under the same set of assumptions. In simulations, we instead find that the negative dilation arrays first buckle at higher values of $\gamma$ ($\gamma_c=26.1$) compared to positive dilation arrays ($\gamma_c=20.8$). We previously showed that the nonlinear continuum theory introduced in ref. \cite{plummer2020buckling} is able to capture this delay \cite{hanakata2022anomalous}. 

As before, away from the buckling threshold, we can numerically minimize the system with respect to $\eta$ and $f$. When $\epsilon=-0.1, k=100 \hat{\kappa}/{a_0^2}$, we now find that $\eta_{\text{min}}=0.919$ and $f_{\text{min}}=0.346 a_0$. These values are again identical to those measured in simulations of a small system. We note that this $\sim 8\%$ contraction is more significant than the $\sim 2\%$ expansion found for positive dilation arrays at the same magnitude of $\gamma$. 

\section{Elimination of tangential phonons}\label{sec:derive}
We present here additional details on how to eliminate the tangential displacements in the energy functional defined by Eqs. \ref{eq:estretch} and \ref{eq:ebend}.

If we wish to work at finite temperature, we can integrate over the tangential phonons in the partition function. At zero temperature, we instead assume the energy is minimized with respect to tangential phonons. For either scenario, the first step in the calculation is to shift $u_\alpha$ by a function of $f$ such that the energy becomes quadratic in the variable containing $u_\alpha$ (completing the square).  

The appropriate shifted variable in Fourier space is 
\begin{align}
    w_\alpha(\*q)&=u_{\alpha}(\mathbf{q})+\phi_\alpha(\*q)-\frac{\lambda}{2 \mu+\lambda} \frac{i q_{\alpha}}{q^{2}}\left(\Phi(\mathbf{q})+\frac{f(\mathbf{q})}{R}\right)\nonumber \\&+\frac{(\mu+\lambda)}{2 \mu+\lambda} \frac{i q_{\alpha}}{q^{2}}\left(\Omega_{0} c(\mathbf{q})+2 \frac{q_{1}^{2}}{q^{2}} \frac{f(\mathbf{q})}{R}\right)-2 \frac{i q_{1} \delta_{1 \alpha}}{q^{2}} \frac{f(\mathbf{q})}{R},
\end{align}
with the following notation \cite{plummer2020buckling}: $\phi_\alpha(\*q)$and $\Phi(\*q)$ are the longitudinal and transverse parts of $A_{\alpha \beta}(\*q)$ respectively, where $A_{\alpha \beta}(\*r)= \frac{1}{2} \left(\frac{\partial f}{\partial x_{\alpha}} \frac{\partial f}{\partial x_\beta} \right)$, and $c(\*q)$ is the Fourier transform of the concentration of dilations, $c(\*r)= \sum_i \delta^2(\*r-\*r_i)$. The Fourier convention used is $f(\*r)=\sum_{\*q} f(\*q) e^{i\*q \cdot \*r}$, $f(\*q)=\frac{1}{A} \int d^2 r f(\*r)e^{-i\*q \cdot \*r}$. 

In terms of the shift variable $w_\alpha(\*q)$, the stretching energy (Eq. \ref{eq:estretch}) becomes
\begin{align}
    \frac{E}{A}&=\mu \left(\tilde{u}_{\alpha \beta}^0\right)^2 + \frac{\lambda}{2} \left(\tilde{u}_{\gamma \gamma}^0\right)^2 +  \frac{\kappa}{2 R^2}+ \frac{\kappa}{2}\sum_{\*q}q^4 |f(\*q)|^2
    \nonumber \\ &+\sum_{\*q \neq 0} \left(\frac{\mu}{2}q^2|\*w(\*q)|^2+ \frac{\mu+\lambda}{2}|\*q \cdot \*w(\*q)|^2\right)\nonumber \\ &+\frac{Y}{2}\sum_{\*q\neq 0}\left| \Phi(\*q) - \frac{\Omega_0}{2} c(\*q)+P_{11}^T(\*q) \frac{f(\*q)}{R} \right|^2,
\end{align}
where $Y= \frac{4 \mu(\mu+\lambda)}{2\mu+\lambda}$ is the 2D Young's modulus, $A$ is the area of the system, and $P_{11}^T=1-\frac{q_1^2}{q^2}$. 

Upon minimizing (or integrating) over $w_\alpha(\*q)$ and $\tilde{u}_{\alpha \beta}^0$, we find the energy (or free energy) as a function of $f$: 
\begin{equation}
\small
    E=\frac{\kappa A}{2 R^2}+\frac{A}{2}\sum_{\*q\neq 0} \left[\kappa q^4 |f(\*q)|^2+ Y\left| \Phi(\*q) - \frac{\Omega_0}{2} c(\*q)+P_{11}^T(\*q) \frac{f(\*q)}{R} \right|^2\right],
    \label{eq:ftcompact}
\end{equation}

In real space, this functional becomes
\begin{align}
    E=&\frac{\kappa}{2} \int d^{2} r\left(\nabla^{2} f-\frac{1}{R}\right)^{2}\nonumber \\&+\frac{Y}{2} \int^{\prime} d^{2} r\left(\frac{1}{2} P_{\alpha \beta}^{T} \partial_{\alpha} f \partial_{\beta} f-\frac{\Omega_{0}}{2} c(\mathbf{r})+P_{11}^{T} \frac{f}{R}\right)^{2},
\end{align}
where the prime on the second integral signals that the $\*q=0$ mode is excluded.

\section{Molecular dynamics simulations}
\label{sec:md-details}
All simulations are performed on HOOMD-blue package v2.8.1~\cite{anderson2020hoomd}. Simulation details for the planar membranes can be found in our recent work~\cite{hanakata2022anomalous}. We set $k=100 \hat{\kappa}/a_0^2, \hat{\kappa}=1$ and $a_0=1$. Temperatures are reported in units of $\hat{\kappa}$. We vary the temperature from $T=0.05$ to $T=0.500$. Following our previous work~\cite{hanakata2022anomalous}, we initialize the heights of the puckers with the ground state pattern, either an antiferromagnetic (AFM) or ferromagnetic (FM) configuration depending on the radius. A small amount of noise is added to every node. We then perform zero-temperature structural relaxation using the FIRE algorithm~\cite{fire-algo} with force and energy convergence criteria of $10^{-6}$ and $10^{-10}$, respectively, and a step size $dt=0.005$ to minimize energy and stress.

At finite temperature, NPT (fixed number of particles, pressure, temperature) molecular dynamics simulations with zero-stress condition are used after employing the zero-temperature structural relaxation. Pressure and temperature are controlled by the Martyna-Tobias-Klein barostat-thermostat~\cite{martyna1994constant} with a time step $dt=0.001$, thermostat coupling $\tau_T=0.2$ and barostat coupling $\tau_P=1.0$. Periodic boundaries are applied in the $x$ and $y$ directions for the membranes and along the tube axis for the cylinders. NPT simulations are run for $10^7$ time steps for cylinders with $L<60a_0$ and $2\times 10^7$ for cylinders with $L\geq60a_0$. Snapshots are taken every 10,000 steps and the first half of data is discarded for thermal equilibration. We typically perform 10 independent runs at each temperature and 20+ independent runs closer to the transition temperature. HOOMD simulation input scripts and other codes are available at \url{https://github.com/phanakata/programmable-matter} 

\subsection{Thermal equilibration with random initial conditions}
\begin{figure}
\begin{center}
\includegraphics[width=\linewidth]{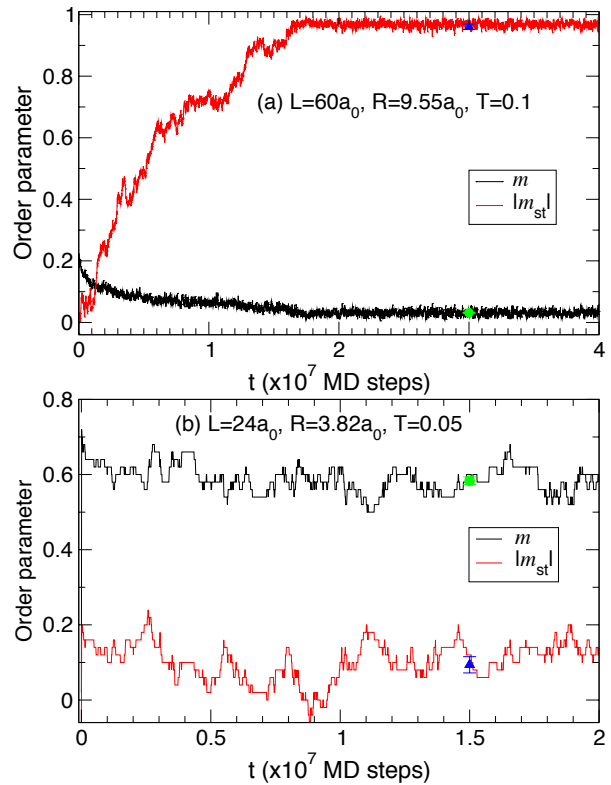}
\end{center}
\caption{{\label{fig:random} Magnetization $m$ and staggered magnetization  $\mst$ as a function of time $t$ of cylinders with (a) $L=60a_0$ ($R>R_t$) and (b) $L=24a_0$ ($R<R_t$) prepared using random initial conditions. At $t=0$, both $m$ and $\mst$ are approximately zero. At long times, $m$ and $\mst$ are close to the average values of systems prepared with ground state initial conditions (green circle and blue triangle, respectively).}}
\end{figure} 
 To check the robustness of our thermalization protocol, Fig. \ref{fig:random} shows additional simulations of systems prepared with random initial conditions. Specifically, we initialize the pucker heights of cylinders with $L=60a_0$ ($R>R_t$, AFM ground state) and $L=24a_0$ ($R<R_t$, FM ground state) with random values and omit the zero-temperature structural optimization. We find that the order parameters converge to the average values obtained from systems prepared with ground state initial conditions but in a much longer time. We therefore save computational resources by performing simulations with ground state initial conditions as described above.

\subsection{Extending axial length}
\label{sec:axial}
\begin{figure}
\begin{center}
\includegraphics[width=\linewidth]{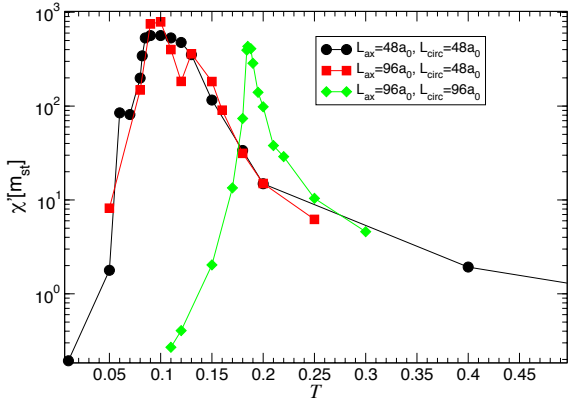}
\end{center}
\caption{{\label{fig:diffaxial}  Staggered susceptibility $\chi'[\mst]$ as a function of temperature $T$ for three cylinders, $[(L_{\rm ax}=48a_0, L_{\rm circ}=48a_0), (L_{\rm ax}=96a_0, L_{\rm circ}=48a_0), (L_{\rm ax}=96a_0, L_{\rm circ}=96a_0)]$. Doubling the axial length $L_{\rm ax}$ while keeping the curvature ($L_{\rm circ}$) fixed has small effect on the effective $T_c$ (the location of the peak in $\chi'[\mst]$).}}
\end{figure} 
Here we provide additional simulations of a cylinder in which we vary both the axial length, $L_{\rm ax}$ and the circumference, $L_{\rm circ}$. In the main text, these lengths were always equal. In Fig. \ref{fig:diffaxial}, we observe that the staggered susceptibility of the medium cylinder with $L_{\rm ax}=L_{\rm circ}=48 a_0$ is similar to the staggered susceptibility of a cylinder with $L_{\rm ax}=96 a_0, L_{\rm circ}=48 a_0$. However, doubling the circumference as well leads to a very different behavior, consistent with Fig. \ref{fig:Xi1}.

\section{Ising antiferromagnet in a uniform field}
\label{sec:mft}
Here, we provide a standard derivation for the free energy expansion for a conventional Ising antiferromagnet on a square lattice in a uniform external field using Bragg-Williams mean-field theory (see, e.g., ref. \cite{chaikin}) and provide references on how to improve these results. The resulting Landau-like theory bears some resemblance to the energy functionals we find for puckers on a cylinder (Eqs. \ref{eq:discrete_mmst} and \ref{eq:ssh_mmst}).

The energy of an Ising antiferromagnet with a spin configuration $\{\sigma_j\}$, $\sigma_j= \pm 1$ in the presence of an external field $h$ that favors up spins is
\begin{equation}
E=J\sum_{\langle i,j\rangle}\sigma_i\sigma_j-h\sum_{i}\sigma_i,
\label{eq:isingen}
\end{equation}
with $J>0$. 

The square lattice is bipartite, so it can be divided into sublattices $\mathcal{A}$ and $\mathcal{B}$ such that all interactions are between (and not within) sublattices $\mathcal{A}$ and $\mathcal{B}$. We define $m_A=\frac{1}{N_A}\sum_{i \in \mathcal{A}} \sigma_i$ and $m_B=\frac{1}{N_B}\sum_{i \in \mathcal{B}} \sigma_i$, the uniform magnetization per spin of each sublattice. 

Each sublattice has an entropy of mixing due to the number of ways to achieve a magnetization $m_{A,B}$ with $N_{A,B}$ spins:
\begin{equation}
\small
\begin{aligned}
    s(m_{A,B})=\frac{S(m_{A,B})}{N_{A,B}} \approx&- k_B  \Bigg[ \left(\frac{1+m_{A,B}}{2}\right) \log \left(\frac{1+m_{A,B}}{2} \right) \\&+ \left(\frac{1-m_{A,B}}{2}\right) \log \left(\frac{1-m_{A,B}}{2} \right)\Bigg]. \label{eq: entropy}
\end{aligned}
\end{equation}
We now approximate Eq. \ref{eq:isingen} by replacing each spin with a spatial average. On combining the approximated energy with Eq. \ref{eq: entropy}, the Bragg-Williams free energy per spin for an Ising antiferromagnet reads
\begin{align}
    \frac{F}{N}&=\frac{E-TS}{N}\nonumber \\&=2 J m_A m_B -\frac{h}{2}(m_A + m_B)- \frac{T}{2} \left(s(m_A)+s(m_B)\right).
    \label{eq:mft_en}
\end{align}

In the limit of small $h$, with $T$ close to $T_c$, we expect both $m_A$ and $m_B$ to be small. We now expand in $m_A$ and $m_B$, neglecting terms of order $m_{A,B}^5$:
\begin{align}
    \frac{F}{N} \approx &-k_B T \log 2 +2 J m_A m_B -\frac{h}{2}(m_A + m_B)\nonumber \\&+ \frac{k_B T}{4} (m_A^2+m_B^2) + \frac{k_B T}{24} (m_A^4+m_B^4).
\end{align}

We then define the magnetization and staggered magnetization in terms of $m_A$ and $m_B$:
\begin{align*}
    m&=\frac{1}{2}(m_A+m_B),\\
    \mst&=\frac{1}{2}(m_A-m_B).
\end{align*}

Upon making these substitutions, the expansion becomes
\begin{align}
    \frac{F}{N}\approx &-k_B T \log 2 - h m +2J (m^2 -\mst^2)+ \frac{k_B T}{2}(m^2+ \mst^2)\nonumber \\&+ \frac{k_B T}{12}( m^4+\mst^4) + \frac{k_B T}{2}m^2 \mst^2 .
    \label{eq:mft_all}
\end{align}

We see that the coefficient of the term quadratic in $\mst$ changes sign at $k_B T_c=\frac{4 J}{1+m^2}\approx 4J -4J m^2$. The coupling with $m$ shifts the transition temperature but does not affect the nature of the phase transition. When $h=0$, $m=0$, and we regain the mean-field theory critical temperature of the Ising model, $k_B T_c^0=4J$. Upon neglecting terms quartic in $m$ and quadratic in $\mst$, we minimize $f$ with respect to $m$ and estimate $m\approx \frac{h}{4J+k_B T}$. On substituting this result in our estimate of $T_c$, we find the shift in the critical temperature as a function of the external field $k_B T_c \approx 4 J - 4 J \left( \frac{h^2}{64 J^2} \right)= 4J- \frac{h^2}{16J}$. Unlike the ferromagnetic Ising model, we have a critical line in the $h-T$ plane, rather than a single critical point. With the mean field approximations described above, the critical line is given by
\begin{equation}
   h(T_c)=\pm \sqrt{16J k_B (T_c^0-T_c)}. 
\end{equation}

Many better approximations for the critical line have been derived \cite{blote1990accurate,wang1997critical,tarasenko1999adatom,monroe2001systematic}. The ``interface solution" of \citet{muller1977interface}, though not the most accurate among them \cite{baxter1980hard}, has a particularly simple form and agrees well with simulations \cite{binder1980phase} and exact results in the limits $h \to 0$ and $T \to 0$. 
\begin{equation}
    \cosh\left(\frac{h}{k_B T_c}\right)=\sinh^2\left(\frac{2 J}{k_B T_c}\right).
\end{equation}

As discussed in the main text, we observe that Eq. \ref{eq:mft_all} has many of the same terms as Eqs. \ref{eq:discrete_mmst} and \ref{eq:ssh_mmst}--a linear term coupling the field and magnetization, and all even terms in $m$ and $\mst$. However, Eq. \ref{eq:mft_all} is missing terms that scale as $m^3$ and $m \mst^2$. Terms of this form would be created if Eq. \ref{eq:mft_en} had a term $w(m_A^3+m_B^3)=2w m^3+6w m \mst^2$. 

%

\end{document}